\documentclass[12pt]{iopart}
\pdfoutput=1
\usepackage{graphicx}
\usepackage{amstext}
\usepackage{makeidx}
\usepackage{amssymb,amsfonts}
\usepackage{verbatim}
\usepackage{epstopdf}
\usepackage{hyperref} 
\usepackage{color}
\usepackage{bm}

\def\P{{\mathbb P}}

\def\res{\textrm{res}}

\begin{document}

\title{A model of non-Gaussian diffusion in heterogeneous media}
\author{Yann Lanoisel\'ee}
  \ead{yann.lanoiselee@polytechnique.edu}
\address{Laboratoire de Physique de la Mati\`{e}re Condens\'{e}e (UMR 7643), \\ 
CNRS -- Ecole Polytechnique, University Paris-Saclay, 91128 Palaiseau, France}
\author{Denis S. Grebenkov}
 \ead{denis.grebenkov@polytechnique.edu}
\address{Laboratoire de Physique de la Mati\`ere Condens\'ee, \\
CNRS -- Ecole Polytechnique, F-91128 Palaiseau, France \\ and } 
\address{Interdisciplinary Scientific Center Poncelet (UMI 2615 CNRS/ IUM/ IITP RAS/ Steklov MI RAS/ Skoltech/ HSE), Bolshoy Vlasyevskiy Pereulok 11, 119002 Moscow, Russia}%


\begin{abstract}
Recent progresses in single particle tracking have shown evidences of non-Gaussian distribution of displacements in living cells, both near the cellular membrane and inside the cytoskeleton. A similar behavior has also been observed in granular materials, turbulent flows, gels, and colloidal suspensions, suggesting that this is a general feature of diffusion in complex media. A possible interpretation of this phenomenon is that a tracer explores a medium with spatio-temporal fluctuations which result in local changes of diffusivity. We propose and investigate an ergodic, easily interpretable model, which implements the concept of diffusing diffusivity. Depending on the parameters, the distribution of displacements can be either flat or peaked at small displacements with an exponential tail at large displacements. We show that the distribution converges slowly to a Gaussian one. We calculate statistical properties, derive the asymptotic behavior, and discuss some implications and extensions. 
\end{abstract}



\date{\today}
\maketitle

\section{Introduction}

A reliable description of transport processes in complex media, such as living cells, is a challenging problem. The primary biological motivation is to understand how the intracellular transport can be efficient enough to allow cell life in crowded environments. From the physical perspective, the challenge stands in elaborating a unified mesoscopic description of transport in disordered media which is consistent with experimental observations of single particle trajectories.

The past years witnessed numerous experimental observations of anomalous diffusion with the Mean Squared Displacement (MSD) evolving with time as a power law $\langle X^2(t)\rangle\propto t^\alpha$ \cite{Weiss2004,Barkai2012,Bertseva2012,Hofling2013,Manzo2015,Sadegh2017}. Many models have been proposed to rationalize this power law \cite{Bouchaud1990,Havlin2002,Metzler2014,Spanner2016}, each model providing a different interpretation of the effect of crowding on tracer's motion in the subdiffusive case $\alpha<1$. The anomalous scaling can originate from (i) diffusion in a fractal medium due to the excluded volume; (ii) viscoelastic properties of the medium described by fractional Brownian motion \cite{Mandelbrot1968} or generalized Langevin equation \cite{Wang1992,Porra1996,Grebenkov2011b}; (iii) molecular caging when the tracer is stopped for a random power law distributed time described by Continuous Time Random Walk \cite{Montroll1965,Metzler2000}. Having very different physical origins, these models presuppose that dynamic properties of the medium are homogeneous.

Additionally to the anomalous scaling, recent progress in single particle tracking techniques led to the discovery of a class of systems in which individual particles exhibit non-Gaussian diffusion with exponential tails. This interesting feature is not exclusive to microbiology but has appeared in various complex media. Three typical shapes of distribution of displacements have been observed: (i) flat distribution near zero with an exponential tail is found in granular  materials \cite{Orpe2007}, turbulent flow \cite{Gollub1991}, cytoskeleton \cite{Stuhrmann2012}, active gels \cite{Toyota2011,Bertrand2012,Moschakis2012}, glassy material \cite{Chaudhuri2007} and intracellular medium \cite{Grady2017}, (ii) exponential behavior in entangled F-actin networks \cite{Wang2009,Wang2012}, log-return of stock prices \cite{Dragulescu2011}, and (iii) stretched exponential form in granular gas \cite{Rouyer2000}, cell membrane \cite{He2016}, in crowded environments \cite{Ghosh2015,Ghosh2016} and from simulations of diffusion with interacting obstacles \cite{Sarnanta2016}. A common feature of these dynamics is that the displacement distribution becomes Gaussian in the long-time limit \cite{Metzler2017a}.

The abundance of empirical observations suggests that exponential tails are reminiscent of heterogeneous complex media despite different experimental and microscopic setups. Aiming at modeling these exponential tails, we proceed by a mesoscopic approach, which describes the system with time-dependent macroscopic quantities. In this article we focus on the case when non-Gaussian diffusion originates from local changes in diffusive properties of the medium.

Former contributions from the theoretical side started with the K\"{a}rger model \cite{Karger1985,Fieremans2010} in which a particle randomly switches between a finite number of states with different diffusivities. Chubynsky and Slater \cite{Chubynsky2014} modeled diffusivity as a continuous random process with a stationary distribution, and deduced from it the short-time exponential behavior \cite{Wang2009} using superstatistical description \cite{Beck2003,Beck2005}. Jain and Sebastian solved the time-dependent problem in the case when diffusivity is the square of a multidimensional Ornstein-Uhlenbeck process and showed that it can be mapped to a first passage problem in a medium with absorbing sinks \cite{Jain2016,Jain2017b}. They also generalized the solution to the case of L\'{e}vy driven noise \cite{Jain2017a}. Chechkin {\it et al.} \cite{Chechkin2016} solved the problem using subordination technique and pointed out that  superstatistical description matches diffusing diffusivity but only at short times as it cannot reproduce convergence to a Gaussian distribution at long times.\\

We present a three parameter model of non-Gaussian diffusion in which diffusivity is fluctuating around an average value $\bar{D}$ $(m^2/s)$ (which constitutes the effective diffusion coefficient at long time), with the correlation time $\tau$ $(s)$ and the amplitude of fluctuations $\sigma$ $(m/s)$. The description is formulated in terms of coupled Langevin equations from which the characteristic function of displacements is derived in an exact explicit form. The shape of the distribution is tuned by one dimensionless parameter 
\begin{equation}\label{eq:nu}
\nu=\frac{\bar{D}}{\sigma^2\tau},
\end{equation} 
which compares the diffusivity correlation time $\tau$ and the diffusivity fluctuation time $\bar{D}/\sigma^2$. Depending on $\nu$, the distribution of displacements can be close to exponential ($\nu=1$), parabolic ($\nu>1$) or peaked ($\nu<1$) at the origin. In all cases the distribution of displacements exhibits an exponential tail with eventual power law corrections. We show that this description leads to a linear dependence of the MSD on time, while fluctuations of time-averaged MSD span up at long times depicting the effect of heterogeneous diffusivity. 
 We analyze the autocorrelation of squared increments which describes memory loss of diffusivity. These correlations lead to a slow, $1/t$, convergence of the distribution to a Gaussian one. Analytical results are verified numerically by Monte Carlo simulations using Milstein scheme \cite{Higham2001}.
Finally we derive the asymptotic behavior and discuss some implications and generalizations.

\section{Model of non-Gaussian diffusion}
\label{sec:Model_non_Gaussian}
We propose a model of a tracer motion in a heterogeneous medium, in which the diffusivity is a stochastic process instead of being a constant. In order to justify this description, let us consider a single particle tracking measurement of duration $t_{exp}$ with a timestep $\Delta t$ between two position recordings. If the motion occurs in a homogeneous environment, the distribution of displacements becomes Gaussian very fast, in a time $t_{loc}$ of equilibration of the tracer with its local environment. 
For a heterogeneous medium, in which the diffusivity can vary spatio-temporally (noted $D_{x,t}$), we introduce the time $t_{sys}$ for a particle to explore the whole medium and to average diffusivities experienced in the medium. On one hand, if $t_{loc}\ll t_{sys}<\Delta t$, increments of the motion are already coarse-grained at a measurement timestep $\Delta t$ and therefore are Gaussian. On the other hand, if $t_{loc}<\Delta t\ll t_{sys}$, the motion is not necessarily Gaussian because diffusivity evolves in time, and the tracer continuously moves from one equilibrium state to another.
 This can be interpreted as the effect of spatio-temporal heterogeneities in the medium seen from the point of view of a single particle. To simplify the analysis we describe diffusivity as a stochastic process in time, $D_t$, with the idea that the stochasticity is an annealed simplification of the spatio-temporal disorder. The particle experiences a fluctuating diffusivity around an average value $\bar{D}$ toward which the time-averaged effective diffusion coefficient converges at long times (i.e. $t \gg t_{sys} $).  
Two physical constraints for a fluctuating diffusivity are (i) the distribution of displacements converges to a Gaussian one at long times, so diffusivity should have a stationary distribution in the long-time limit, with the average value $\bar{D}$; (ii) diffusivity as a measure of local kinetic energy of the tracer should be non-negative.
 
We propose to model time-dependent diffusivity $D_t$ as a Feller process \cite{Feller1951,Gan2015} or square root process, also known in financial litterature as the  Cox-Ingersoll-Ross process (CIR) \cite{Cox1985}.
This process has been developed in order to rationalize fluctuations of volatility in price asset returns. In the CIR model, the diffusivity fluctuates in a harmonic potential centered on $\bar{D}$ and remains non-negative thanks to the balance between the pulling of harmonic potential and the noise reduction of diffusivity-dependent fluctuations at small $D_t$. Moreover, the stationary distribution of diffusivity is known to be a Gamma distribution.
For the sake of clarity, we first formulate the model for one-dimensional motion and then show its straightforward extension to the multi-dimensional isotropic case. For a tracer starting at $x_0$ with diffusivity $D_0$, the corresponding coupled Langevin equations read:
\begin{equation}\label{Langevin_Equation}
\left\{
    \begin{array}{ll}
dx_t=\sqrt{2D_t}d W_t^{(1)}, \\
dD_t=\frac{1}{\tau}(\bar{D}-D_t)dt+\sigma\sqrt{2D_t}dW_t^{(2)},
    \end{array}
\right.
\end{equation}
where $x_t$ and $D_t$ are stochastic time-dependent position and diffusivity of the tracer, $dW_t^{(1)}$ and $dW_t^{(2)}$ are increments of independent Wiener processes (white noises). The model includes three parameters: the average diffusivity $\bar{D}$ (in m$^2$/s), the correlation time $\tau$ (in s) and the amplitude of fluctuations $\sigma$ (in m/s).

The approach by Chubynsky and Slater \cite{Chubynsky2014} is retrieved by setting a diffusivity bias $s(D)=-\frac{1}{\tau}(D-\bar{D})$ and a diffusivity of diffusivity $d(D)=\sigma^2\sqrt{2D}$, although in our model, a reflecting boundary is necessary only at $D=0$ (see below). Jain and Sebastian \cite{Jain2016}
and Chechkin {\it et al.} \cite{Chechkin2016} considered the diffusivity as the distance from the origin of an $n$-dimensional Ornstein-Uhlenbeck process, which is a particular case of our model. In \ref{sec:BDeriv_CIR} we present the derivation of the Cox-Ingersoll-Ross model starting from the $n$-dimensional Ornstein-Uhlenbeck process (note that the relation between these models was already mentioned in \cite{Chechkin2016}). It is then evident that the results of former studies can be reproduced for integer values $n=\frac{\bar{D}}{\sigma^2\tau}$ and the range of applicability is thus widened because parameters in our model are continuous: $\lbrace \tau,\bar{D},\sigma\rbrace\in(0,\infty)$. In particular, the case $\nu<1$ (see Eq. \ref{eq:nu}), which yields a peaked distribution of displacements and the most peculiar properties of heterogeneous diffusion (see below), was not accessible so far.  

We introduce the propagator $P(x,D,t\vert x_0,D_0)$, the probability density for a tracer to be at $x$ with diffusivity $D$ at time $t$, when started from $x_0,D_0$ at $t=0$. The corresponding forward Fokker-Planck equation in the It\^o interpretation reads
\begin{equation}\label{FP_xD}
\frac{\partial}{\partial t}P(x,D,t\vert x_0,D_0)=\frac{1}{\tau}\frac{\partial}{\partial D}\left[(D-\bar{D})P\right]+\frac{\partial^2}{\partial x^2}\left(DP\right)+\sigma^2\frac{\partial^2}{\partial D^2}\left(DP\right),
\end{equation}
with the initial condition $P(x,D,t=0\vert x_0,D_0)=\delta(x-x_0)\delta(D-D_0)$.\\
Following Dr\v{a}gulescu and Yakovenko \cite{Dragulescu2011}, this equation is solved by performing the Fourier transform with respect to position $x$, and the Laplace transform with respect to diffusivity $D\geq 0$:
\begin{equation}\label{FLTrans}
\tilde{P}(q,s,t\vert x_0,D_0)=\int_{-\infty}^\infty dx\int_0^\infty dDe^{-iqx-Ds}P(x,D,t\vert x_0,D_0),
\end{equation}
where $q$ and $s$ are the dual variables to position and diffusivity, respectively. Inserting Eq. (\ref{FLTrans}) into Eq. (\ref{FP_xD}) leads to the first order partial differential equation:
\begin{equation}\label{Eq_Gen_TF_TL}
\frac{\partial}{\partial t}\tilde{P}+\left(\sigma^2s^2+\frac{1}{\tau}s-q^2\right)\frac{\partial}{\partial s}\tilde{P}=-\frac{1}{\tau}\bar{D} s\tilde{P}
+J_D(D=0,t),
\end{equation}
subject to the initial condition $\tilde{P}(q,s,t=0|x_0,D_0)=e^{-iqx_0}e^{-sD_0}$. The last term $J_D(D=0,t)=\left(\frac{\bar{D}}{\tau}-\sigma^2\right)P(q,D=0,t\vert X_0,D_0)$ can be interpreted as a probability density flux across the boundary $D=0$ in the phase space $(x,D)$. In the case of an {\it absorbing} diffusivity boundary at $D=0$, there is an atom of probability measure at $D=0$ which ``accumulates'' absorbed trajectories. The probability of having $D=0$ grows with time and is related to the function $J_D(D=0,t)$ which can be deduced from initial conditions by solving an integral equation (see \cite{Feller1951} and \ref{sec:solution_absorbing_boundary}). In our model, the diffusivity cannot be physically zero and the distribution must have a nontrivial stationary solution to match convergence to a Gaussian distribution at long times, so we choose {\it reflecting} boundary condition at $D=0$ and thus $J_D(D=0,t)=0$. In this case, the solution for any $\nu\geq 0$ is (see \ref{sec:sol_reflect_bound} for detailed derivation)
\begin{eqnarray}\label{eq:complete_solution}
\nonumber
\fl
\tilde{P}(q,s,t\vert x_0,D_0)&=&
F(x_0,D_0,s)\\
\fl
&&\times\left(\frac{\sigma^2\tau}{\omega}\left[\left(s+\frac{1+\omega}{2\sigma^2\tau}\right)-\left(s+\frac{1-\omega}{2\sigma^2\tau}\right) e^{-\omega t/\tau}\right]e^{ \left(\frac{\omega-1}{2}\right) t/\tau}\right)^{-\nu},
\end{eqnarray}

with $F(x_0,D_0,s)=\exp\left[-iqx_0-\frac{D_0}{2\sigma^2\tau}\left(-1-\omega+\omega\frac{2}{1-\xi e^{-\omega t/\tau}}\right)\right]$, $\xi=1-\frac{2\omega}{1+\omega+2\sigma^2\tau s}$ and $\omega=\sqrt{1+4\sigma^2\tau^2q^2}$.
The inverse Fourier and Laplace transforms yield $P(x,D,t|x_0,D_0)$. However, this solution provides too detailed information which can hardly be confronted to single particle tracking data with no direct access to diffusivities $D$ and $D_0$. We thus integrate the solution over $D$ (which is equivalent to set $s=0$) to get the marginal distribution of positions. We also assume that the tracer's initial diffusivity $D_0$ is taken from its stationary Gamma distribution $\Pi(D_0)$ (see \ref{sec:sol_reflect_bound}):
\begin{equation}\label{eq:stat_CIR}
\Pi(D_0)=\frac{\nu^\nu D_0^{\nu-1}}{\Gamma\left(\nu\right)\bar{D}^\nu}\exp\left(-\frac{\nu}{\bar{D}}D_0\right),
\end{equation}
where the shape parameter $\nu$ is defined in Eq. (\ref{eq:nu}).
 The average over $D_0$ yields the marginal distribution 
\begin{equation}\label{eq:marginal_solution}
 P(x,t\vert x_0)=\frac{1}{2\pi}\int_{-\infty}^\infty dq\: e^{iq(x-x_0)}\tilde{P}(q,t),
\end{equation} 
 with
\begin{equation}\label{eq:margin_sol_no_init}
\tilde{P}(q,t)=\left(e^{-\frac{1}{2}\left(\omega-1\right) t/\tau}
\frac{4\omega}{\left(\omega+1\right)^2}\left(1-\left(\frac{\omega-1}{\omega+1}\right)^2
e^{-\omega t/\tau} \right)^{-1}
\right)^\nu.
\end{equation} 
An alternative solution using the subordination concept, inspired from \cite{Chechkin2016}, is given in \ref{sec:Subordination}.\\
When particles undergo isotropic motion in $\mathbb{R}^d$, the formula for the distribution of displacements remains almost unchanged, except that one has to perform multi-dimensional inverse Fourier transform in $\mathbb{R}^d$, with ${\bf q},{\bf x}$ and ${\bf x_0}$ being vectors:
\begin{equation}
P({\bf x},t\vert {\bf x_0})=\int\limits_{\mathbb{R}^d} \frac{d^d{\bf q}}{(2\pi)^d}\: e^{i {\bf q}({\bf x}-{\bf x_0})}\tilde{P}(|{\bf q}|,t),
\end{equation}
with $w=\sqrt{1 + 4\sigma^2\tau^2|{\bf q}|^2 }$. 
Since the characteristic function $\tilde{P}(|{\bf q}|,t)$ depends only on $|{\bf q}|$, one can use spherical coordinates and integrate out the angular variables, yielding
\begin{equation}
P(r,t)=\frac{r^{1-d/2}}{(2\pi)^{d/2}}\int\limits_0^\infty dq\: q^{d/2}J_{\frac{d-2}{2}}(qr)\:\tilde{P}(q,t),
\end{equation}
where $J_{\alpha}(x)$ is the Bessel function of the first kind, $r=|{\bf x}-{\bf x_0}|$, and $\tilde{P}(q,t)$ is given by Eq. (\ref{eq:margin_sol_no_init}). In what follows, we focus on the one-dimensional case, bearing in mind straightforward extensions to the multi-dimensional case.

\begin{figure*}[h!]
\begin{center}  
\includegraphics[width=140mm]{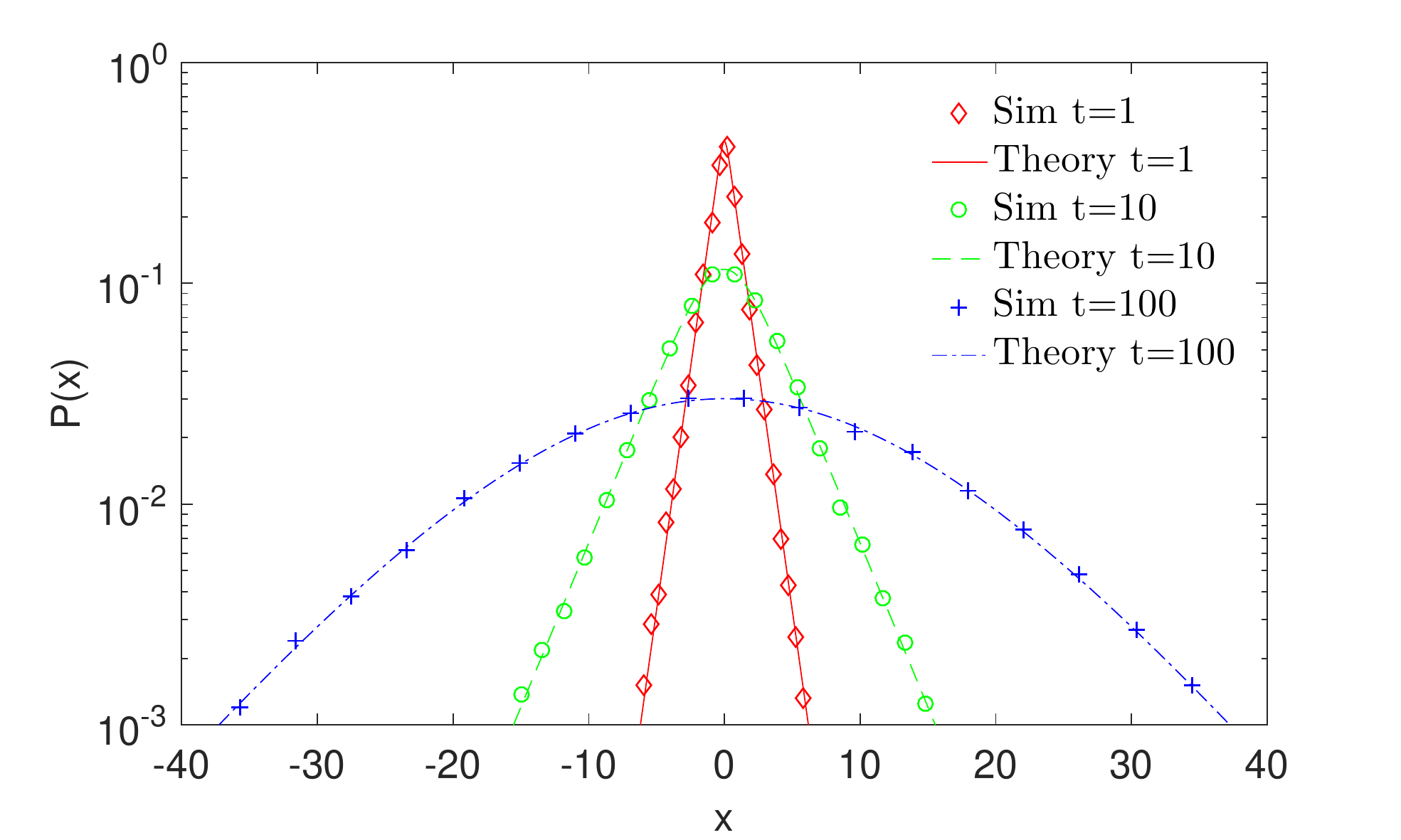}\end{center}
\caption{Distribution of displacements at times $t=\lbrace 1,10,100\rbrace$. Here $\tau=10$, $\bar{D}=1$, $\sigma=1/\sqrt{\tau}$ and thus $\nu=1$. Theoretical results (lines) are compared to Monte Carlo simulations (symbols) with $M=10^6$ particles.}
\label{fig:Distrib_inc_fct_t}
\end{figure*}

Figure \ref{fig:Distrib_inc_fct_t} illustrates the convergence of the distribution of displacements to a Gaussian one as $t$ increases. Note also a perfect agreement between the theoretical formula (\ref{eq:marginal_solution}) and Monte Carlo simulations.
 
\begin{figure*}[h!]
\begin{center}  
\includegraphics[width=140mm]{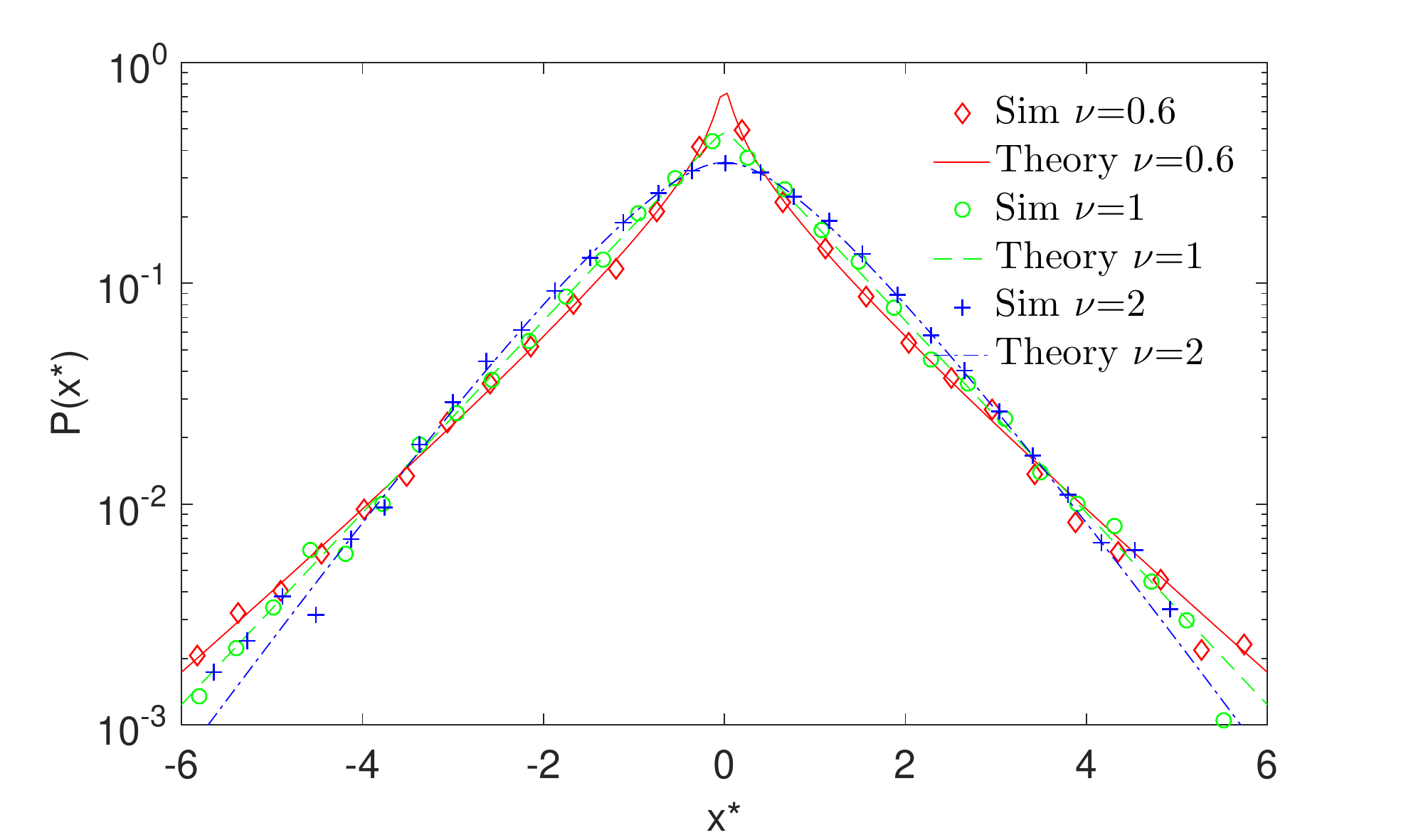}\end{center}
\caption{Distribution of normalized displacements, with $x^*=x/\sqrt{\bar{D}t}$, at fixed time $t=1$ for different parameters $\nu=\lbrace 0.6,1,2\rbrace$. For each case, we kept $\bar{D}=1$ and $\tau=100$ and varied $\sigma$. Theoretical results (lines) are compared to Monte Carlo simulations (symbols) with $M=10^6$ particles.}
\label{fig:Distrib_inc_fct_nu}
\end{figure*}

Figure \ref{fig:Distrib_inc_fct_nu} shows the effect of the shape parameter $\nu$ on the distribution of displacements at time $t=1$. The parameter $\nu$ changes the shape of the distribution. When $\nu\leq 1$, fluctuations are strong compared to both the average diffusivity $\bar{D}$ and the correlation time $\tau$. In this case, the probability of diffusivity close to zero is large that makes the distribution of displacements peaked near $x=0$. In turn, the distribution gets closer and closer to Gaussian as $\nu\rightarrow \infty$ (see Sec. \ref{sec:asymp_large_nu}). 

\begin{figure*}[h!]
\begin{center}  
\includegraphics[width=140mm]{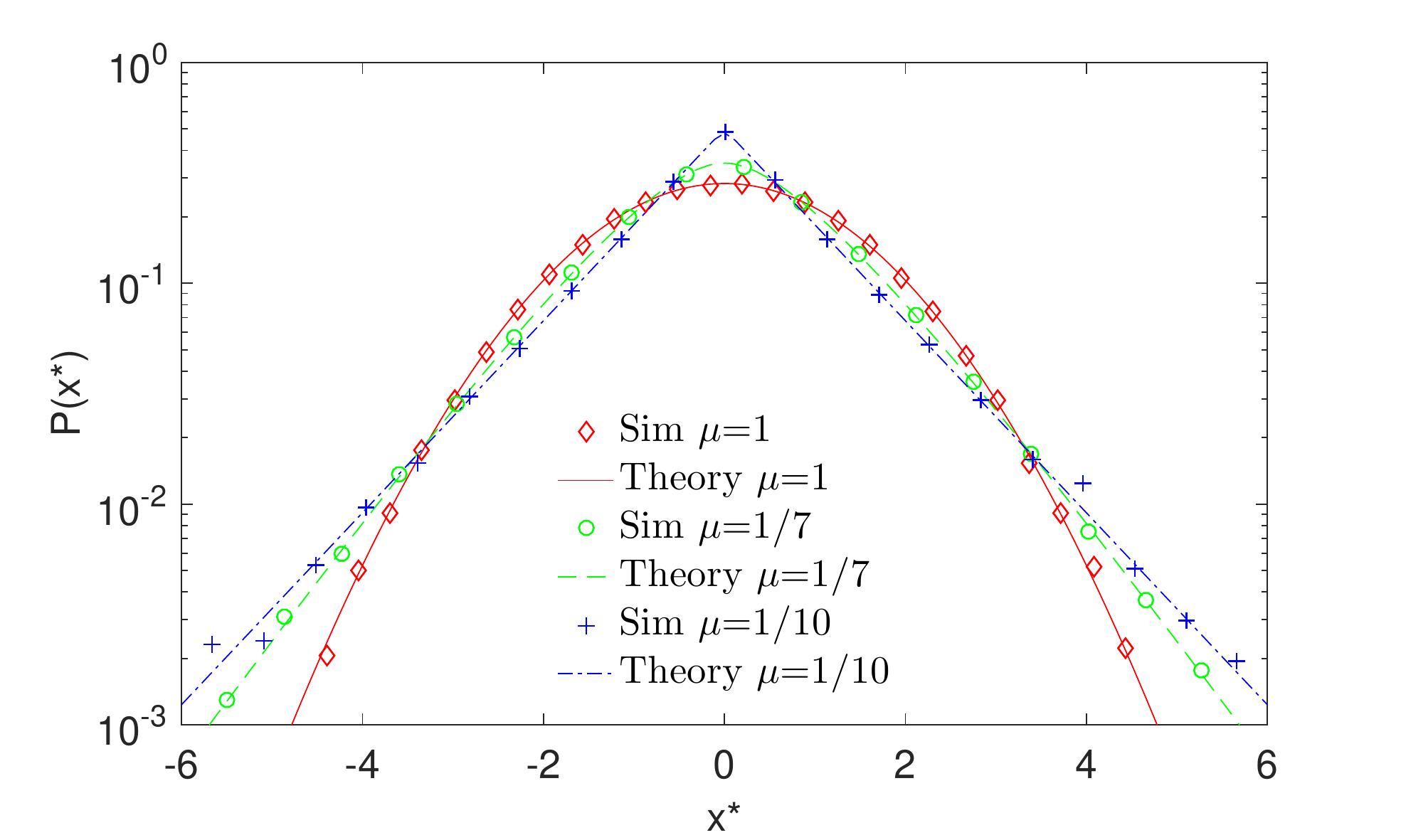}
\end{center}
\caption{Distribution of normalized displacements, with $x^*=x/\sqrt{\bar{D}t}$, at fixed time $t=1$, with $\mu(t)$ from Eq. (\ref{eq_mu}) being varied in the range $\lbrace 1/10,1/7,1 \rbrace$ corresponding to $\nu=\lbrace 1,2,100 \rbrace$, by changing $\sigma$ and keeping $\bar{D}=1$ and $\tau=100$. Theoretical results (lines) are compared to Monte Carlo simulations (symbols) with $M=10^6$ particles.}
\label{fig:Distrib_inc_fct_mu}
\end{figure*}

On the length scale $\sigma\tau$, the diffusivity remains roughly the same. Intuitively, if $\sqrt{\bar{D}t}\ll\sigma\tau$, a particle has not enough time to explore the system. The distribution $P(x,t|x_0)$ could be considered as a superstatistical description of independent particles with constant but randomly chosen diffusion coefficients (see Sec. \ref{subsec:Superstatistics_Short_Time}). Inversely, when $\sqrt{\bar{D}t}\gg\sigma\tau$, the particle
has enough time to explore the medium and the distribution progressively becomes Gaussian. We introduce thus the time-dependent dimensionless diffusion length: 
\begin{equation}\label{eq_mu}
\mu(t)=\frac{\sqrt{\bar{D}t}}{\sigma\tau}.
\end{equation}
 As $\mu(t)\to \infty$, the particle explores the space beyond the correlation length, and the distribution gets closer to a Gaussian one. We show in Fig. \ref{fig:Distrib_inc_fct_mu} how $\mu(t)$ impacts the shape of the distribution.
For instance at $\mu(t)=1$, the distribution is almost Gaussian. When $\mu(t)$ decreases, the distribution becomes more peaked. Note that the quantity $\mu(t)$ is directly related to the non-Gaussian parameter (see Eq. (\ref{eq:NG_Param}) below). 

\begin{figure*}[h!]
\begin{center}  
\includegraphics[width=75mm]{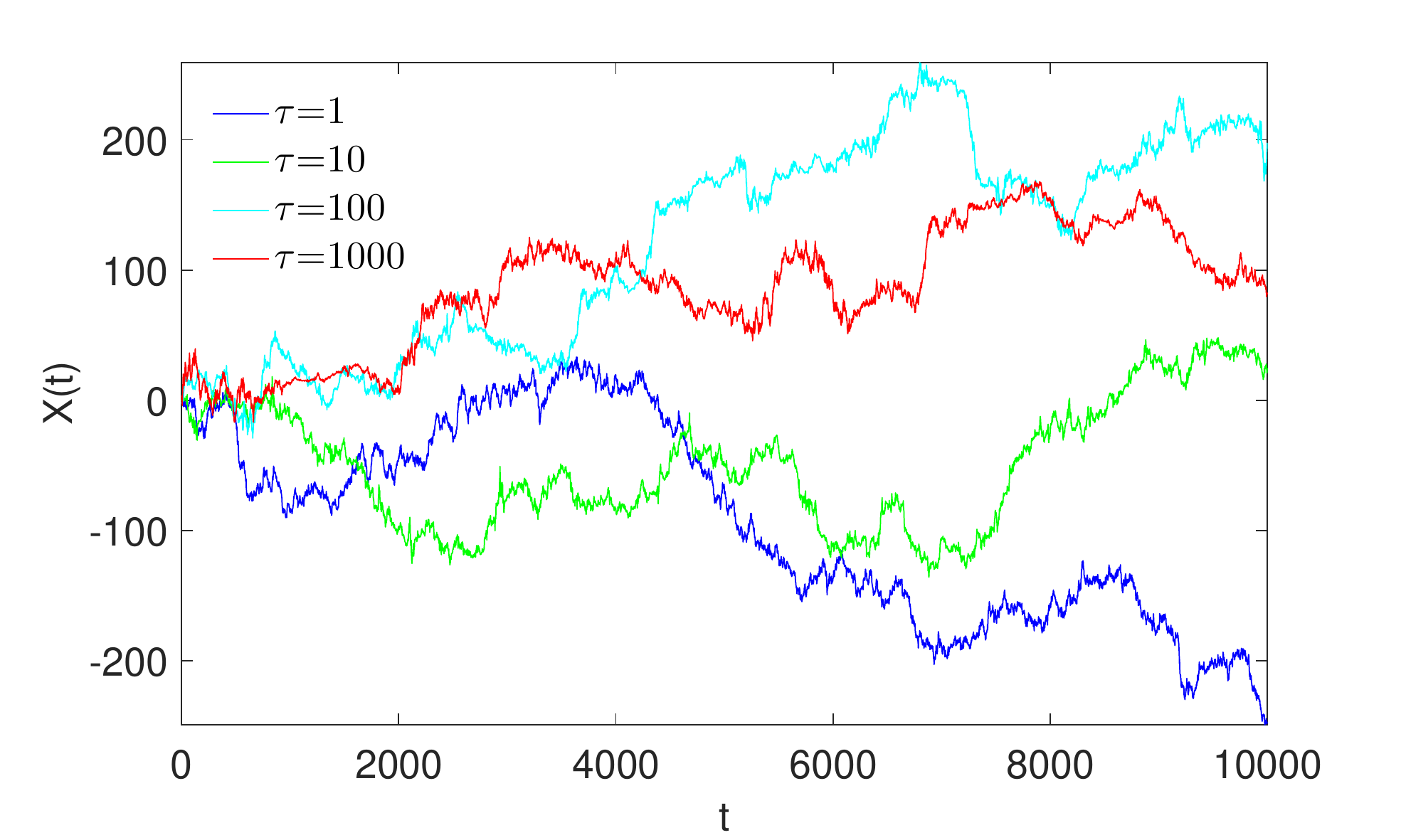}
\includegraphics[width=75mm]
{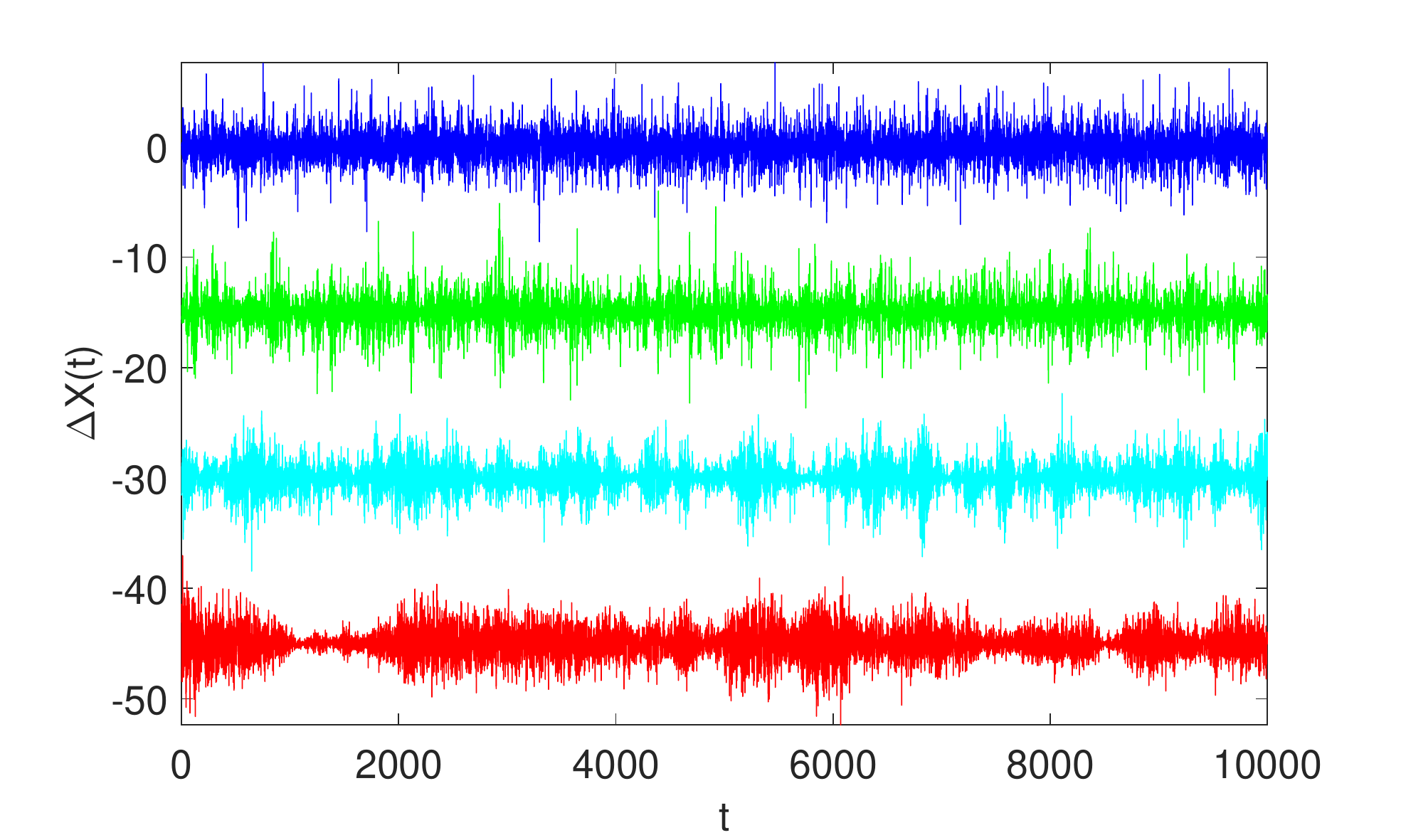}
\end{center}
\caption{\textit{Left.} Trajectories simulated for several values of $\tau=\lbrace 1,10,100,1000\rbrace$. Here $\nu$ is kept equal to one, with $\bar{D}=1$ and $\sigma=1/\sqrt{\tau}$. \textit{Right.} Corresponding time series of position increments with lag-time $\delta t=1$.  For clarity, the time series are artificially shifted (with increasing $\tau$ values from top to bottom), but remain with zero mean.}
\label{fig:TrajInc_var_fct_a}
\end{figure*}

Figure \ref{fig:TrajInc_var_fct_a} illustrates four random trajectories and corresponding displacements. The envelop of time series of displacements shows patterns of fluctuations correlated on timescale $\tau$. For small $\tau$, the envelop becomes constant as in the Brownian motion case.


\section{Asymptotic behavior}
\label{sec:asymptotics}
\subsection{Brownian limit}\label{sec:asymp_large_nu}
We first consider the limiting case $\sigma\tau\rightarrow 0$, which can either be interpreted as diffusivity behaving deterministically ($ \sigma \rightarrow 0$) or 
 the mean reversion significantly stronger than fluctuations of diffusivity ($\tau \rightarrow 0$). In this limit one recovers $\tilde{P}(q,t)=e^{-q^2\bar{D} t}$, from which the Gaussian propagator for Brownian motion is retrieved:
\begin{equation}
P(x,t|x_0)=\frac{1}{\sqrt{4\pi \bar{D} t}}\exp\left(-\frac{(x-x_0)^2}{4 \bar{D} t}\right).
\end{equation}
This distribution also corresponds to the limit $\nu\to\infty$.
\subsection{Short-time behavior}
\label{subsec:Superstatistics_Short_Time}
The superstatistical approach \cite{Beck2003,Beck2005} consists in writing the distribution of displacements as a superposition of Gaussian distributions weighted by a stationary distribution of diffusivity. In a recent work, Chechkin {\it et al.} \cite{Chechkin2016} showed that non-Gaussian diffusion can be described at short times by superstatistics. Since during the correlation time $\tau$, diffusivity does not evolve much, one can imagine an ensemble of particles with independent diffusivities. In our model, we use this relation to establish the short-time behavior.

One can relate our model to the superstatistical approach in the following terms. At short times we have
\begin{equation}
x_t=\int_0^t\sqrt{2D_s}dW_s^{(1)}\approx\sqrt{2D_0}W_t^{(1)},
\end{equation}
and consider $D_0$ in the stationary regime.
 This short-time description looses track of the dynamics. We calculate $P_0(r,t)$, the probability density to be at distance $r$ from the starting point in $d$ dimensions, where the subscript $0$ highlights that it is a short-time description:
 
\begin{equation}
P_0(r,t) =\int_0^\infty dD_0\:\Pi(D_0)\:\frac{1}{(4\pi D_0t)^{d/2}}\exp\left(-\frac{r^2}{4D_0t}\right),
\end{equation}
with the stationary distribution $\Pi(D_0)$ of the CIR model from Eq. (\ref{eq:stat_CIR}), which gives
\begin{equation}
P_0(r,t) =\frac{2^{1-\nu-d/2}\nu^{d/2}}{\Gamma\left(\nu\right)(\pi \bar{D}t)^{d/2}}\left(r\sqrt{\frac{\nu}{\bar{D} t}}\right)^{\nu-d/2}K_{\nu-d/2}\left(r\sqrt{\frac{\nu}{\bar{D}t}}\right),
\end{equation}
where $K_{\alpha}(x)$ is the modified Bessel function of the second kind. Using the small $x$ expansion of $K_{\alpha}(x)$, one gets for $\nu>d/2$
\begin{equation}
P_0(r=0,t) =
 \frac{\Gamma(\nu-d/2)}{\Gamma\left(\nu\right)(4\pi t)^{d/2}}\left(\frac{\nu }{\bar{D}}\right)^{d/2}.
\end{equation}
In the case $\nu=1$ and $d=1$, the distribution is purely exponential
\begin{equation}
P_0(r,t) =\frac{1}{2\sqrt{\bar{D}t}}\exp\left(-\frac{\vert r\vert}{\sqrt{\bar{D}t}}\right),
\end{equation}
(note that in this case the displacement $r$ is distributed over $(-\infty,\infty)$ that explains the extra factor $1/2$).
This approach is applicable at short times ($\mu(t)<1$) but fails at long times because the underlying processes are fundamentally different. 
One can compare our model to this approach by calculating the non-Gaussian parameter
\begin{equation}\label{eq:def_NG_parameter}
\gamma(t)=\frac{1}{3}\frac{\langle X^4(t) \rangle}{\langle X^2(t) \rangle^2}-1,
\end{equation} 
which is equal to the excess kurtosis divided by $3$ (the kurtosis of the Gaussian distribution). By definition, the non-Gaussian parameter is zero for the Gaussian distribution. For superstatistics with $d=1$ and $x_0=0$, the MSD is $\langle x^2(t)\rangle_0=2\bar{D}t$ and the fourth moment $\langle x^4(t)\rangle_0=12t^2\bar{D}^2\frac{\nu+1}{\nu}$, which leads to the non-Gaussian parameter:
\begin{equation}
\gamma_0(t)=\frac{1}{\nu}.
\end{equation}
In contrast to our model (see Sec. \ref{sec:mom_and_NG}), the distribution of displacements $P_0(r,t)$ spreads at all times but does not change its shape: changing time just rescales space coordinates of the distribution. From this argument it is clear that the only way to reproduce convergence to a Gaussian distribution at long times is to make the stationary distribution $\Pi(D_0)$
of diffusivity time-dependent, which does not make sense. This is a branching point among non-Gaussian models, as constant or vanishing non-Gaussian parameter implies different miscroscopic mechanisms. Note that the distinction between interpretations can also be made using the autocorrelation of diffusivity: it is a Dirac delta distribution $\delta(\tau)$ in a superstatistical approach and an exponentially vanishing function in our model (see Sec. \ref{sec:autocorr_squared_inc}). 
\subsection{Large $x$ behavior}
\label{subsec:Asymp_large_x}
We investigate the asymptotic behavior of the propagator at large $|x-x_0| \rightarrow\infty$. Here we only summarize the results, while the derivation is detailed in \ref{Annexe:Asymp_large_x}. 
In the case $\nu=1$, we obtain 
\begin{equation}\label{asymp_large_x_nu1}
P(x,t|x_0) \propto \exp\biggl(-\frac{|x-x_0|  \beta_{t^*} }{2\sigma\tau}\biggr)  \qquad (|x-x_0|\to \infty),
\end{equation}
with $\beta_{t^*} = \sqrt{1 + (4\alpha_{t^*}/t^*)^2}$, $t^*=t/\tau$ and $\alpha_{t^*}$ the smallest positive solution of
\begin{equation}
\alpha_{t^*} \sin \alpha_{t^*} = \frac{t^*}{4} \cos\alpha_{t^*}.
\end{equation}

This agrees with experimental observations of a distribution of displacements with exponential tails \cite{Wang2009,Wang2012,Dragulescu2011}.
When $\nu > 1$  is an integer, one gets
power law corrections to the exponential decay:
\begin{equation}\label{asymp_large_x_nudiff1}
P(x,t|x_0) \propto |x-x_0|^{\nu-1} \exp\biggl(-\frac{|x-x_0| \beta_{t^*}}{2\sigma\tau}\biggr)  \qquad (|x-x_0|\to \infty).
\end{equation}
We expect that the same asymptotic behavior remains valid for
any $\nu > 0$ (even non-integer), although its rigorous demonstration
requires much finer analysis and is beyond the scope of this article. We conclude that the propagator exhibits a universal exponential decay at large increments, whereas the value of $\nu$ determines the power law corrections.

\section{Statistical properties}
In this section we describe the statistical properties of our model.
\subsection{Moments and the non-Gaussian parameter}\label{sec:mom_and_NG}

First we calculate the second and fourth moments using the relation 
\begin{equation}
 \langle X^k(t)\rangle =(-i)^k\left(\frac{\partial^k}{\partial q^k}\tilde{P}(q,t)\right)\bigg\vert_{q=0},
\end{equation} 
where $\langle .\rangle$ denotes the expectation.
The second moment reads
\begin{equation}\label{eq:sec_moment}
\langle X^2(t)\rangle =2\bar{D}t.
\end{equation}
We observe thus the mean squared displacement growing linearly with time, as in the Brownian case. In Sec. \ref{sec:Discussion}, an extension to anomalous diffusion through scaling arguments is proposed. 

\begin{figure*}[h!]
\begin{center}  
\includegraphics[width=140mm]{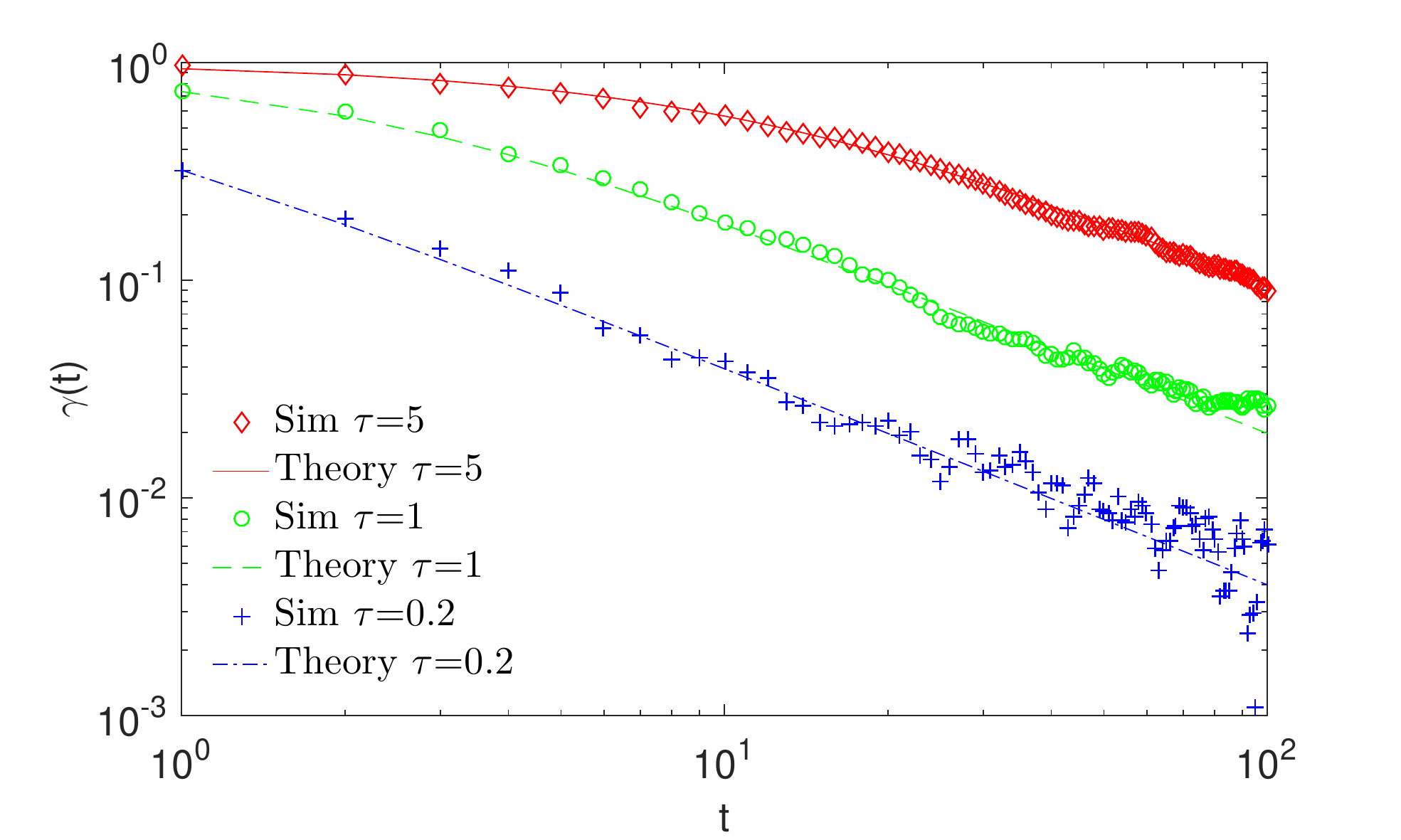}\end{center}
\caption{The non-Gaussian parameter calculated from Eq. (\ref{eq:NG_Param}) (lines) and from Monte Carlo simulations (symbols) with $M=10^5$ particles, for different values of $\tau=\lbrace 0.2,1,5\rbrace$, while keeping $\nu=1$, $\bar{D}=1$ and $\sigma=1/\sqrt{\tau}$.}
\label{fig:Non_Gaussian_parameter}
\end{figure*}

The process described in this article possesses many characteristics which are not deducible from the MSD. So we go further and calculate the fourth moment:
\begin{equation}\label{eq:fourth_moment}
\langle X^4(t)\rangle =12\bar{D}^2 t^2 + 24\sigma^2\bar{D}\tau^2t +24\sigma^2\bar{D}\tau^3\left(e^{-t/\tau}-1\right).
\end{equation}
From the second and fourth moments in Eqs.(\ref{eq:sec_moment},\ref{eq:fourth_moment}), we calculate the non-Gaussian parameter in Eq. (\ref{eq:def_NG_parameter})
\begin{equation}\label{eq:NG_Param}
\gamma(t)=\frac{ 2 \sigma^2\tau^2}{ \bar{D} t} \left(1- \frac{1}{t/\tau}\left( 1-e^{-t/\tau}\right)\right).
\end{equation}
As $t\to\infty$, the distribution slowly converges to a Gaussian distribution, as $1/t$. The theoretical formula is verified by simulations (Fig. \ref{fig:Non_Gaussian_parameter}). The leading term can be expressed in terms of $\mu(t)$ as $\frac{ 2 \sigma^2\tau^2}{ \bar{D} t}=\frac{1}{2}\mu(t)^{-2}$ (see Eq. (\ref{eq_mu})), which shows that non-Gaussianity is related to space exploration, but the complete description also requires to take into account the correction terms from memory effects. Interestingly, we obtained the same form of $\gamma(t)$ as in the K\"{a}rger model \cite{Karger1985,Fieremans2010} with a finite number of equilibrium states (i.e. diffusivities), due to the averaging over diffusivity disorder (see also Sec. \ref{subsec:karger_comp}). The same results are evidently valid for the diffusivity modeled as the distance from the origin of an $n$-dimensional Ornstein-Uhlenbeck process \cite{Jain2017b,Chechkin2016}.

\subsection{Autocorrelation of squared increments}\label{sec:autocorr_squared_inc}
Diffusing diffusivity models introduce a new level of complexity, far beyond the reach of the mean squared displacement analysis, and new tools are needed to describe such processes. A wide range of models with fluctuating volatility (or diffusivity in physical language) have already been studied in finance \cite{Cox1985,Heston1993,Chen1996}. Since the square of an increment is a local measure of diffusivity, its autocorrelations can reveal information on memory effects of diffusivity.
On one hand, it is possible to evaluate the autocorrelation of diffusivity directly from a given trajectory by calculating the autocorrelation of its squared increments. On the other hand, this quantity is accessible theoretically.

Let us define the centered squared increment $dx_t^{2*}=dx_t^2-\langle dx_t^2\rangle$. Generally, one gets (see \ref{sec:Deriv_autoc_squared_inc} for details) 
\begin{equation}
\langle dx_t^{2*}dx_{t+\Delta}^{2*}\rangle=\left\{
    \begin{array}{ll}
12\langle D_t^2\rangle- 4\langle D_t\rangle^2\quad\quad\quad\quad\quad\;(\Delta=0), \\
4\langle D_tD_{t+\Delta}\rangle- 4\langle D_t\rangle\langle D_{t+\Delta}\rangle\quad(\Delta>0). \end{array}
\right.
\end{equation}
In our model, we find
\begin{equation}
\fl
\langle dx_t^{2*}dx_{t+\Delta}^{2*}\rangle=\left\{
    \begin{array}{l}
12\sigma^2\tau\left(1-e^{-t/\tau}\right)^2\left[\bar{D}+2 D_0\frac{e^{-t/\tau}}{1-e^{-t/\tau}}\right]+\\
\quad\quad 8\left((D_0-\bar{D})e^{-t/\tau}+\bar{D}\right)^2\quad\quad\quad\quad\quad\quad\quad\quad\quad\quad\:\:\:\:(\Delta=0),\\
4e^{-\Delta/\tau}\left[\sigma^2\bar{D}\tau\left(1-e^{-t/\tau}\right)^2+2\sigma^2\tau D_0\left(e^{-t/\tau}-e^{-2t/\tau}\right)\right](\Delta>0). \end{array}
\right.
\end{equation}
One notes the exponentially vanishing dependence on initial conditions. In the long-time limit $t\rightarrow \infty$, one simply gets
\begin{equation} 
\lim\limits_{t\to\infty}
\langle dx_t^{2*}dx_{t+\Delta}^{2*}\rangle 
= 
\left\{
    \begin{array}{ll}
12\sigma^2\bar{D}\tau+8\bar{D}^2\quad (\Delta=0), \\
4\sigma^2\bar{D}\tau e^{-\Delta/\tau}\quad\quad\,(\Delta>0).   \end{array}
\right.
\end{equation}
The mean-reverting property of the Feller process results in the exponential autocorrelation of diffusivity.
If an experimentally measured autocorrelation of squared increments is not exponentially vanishing, the mean reverting property cannot be described by a simple harmonic potential centered on $\bar{D}$, and thus another model (or an extension of the present model) should be considered.

\subsection{Ergodicity and finite sample effects}
\label{subsec:Ergodicity}
Data analysis is usually performed with time-averaged quantities because of small data samples. Then a natural question of equivalence between time and ensemble averages arises: ``Is a time-averaged quantity from one particle representative of other particles from the same system?''. For a system at thermodynamical equilibrium, the time average over an infinitely long trajectory matches the ensemble average over an infinite number of particles,  this statement is known as the ergodicity hypothesis. This hypothesis is not satisfied in aging random media \cite{Bouchaud1990}.

From the Langevin equation (\ref{Langevin_Equation}), one can directly see that our model is ergodic: as the diffusivity is fluctuating around its average, fluctuations will be averaged out in the limit of infinitely long trajectories. But for a finite duration of experiment, what can be said about ergodicity of the system? 
 
If the experiment duration $t_{exp}$ is shorter than the time to explore heterogeneties of the system, $t_{exp}< t_{sys}$, different tracers probe regions with different diffusivities. As a consequence, on such a timescale, tracers would appear as experiencing different dynamics, so that one could wrongly conclude that the dynamics of the system is nonergodic. Inversely, if the experiment is sufficiently long (i.e. $t_{exp}\gg t_{sys}$), tracers have enough time to visit every region of the system, and one concludes correctly that the ergodicity hypothesis is fulfilled. The experiment duration plays therefore an important role and should be chosen accurately.

To illustrate this point we study two quantities characterizing ergodicity by different strategies. We show that depending on the parameters of the model, the results of the tests can sound contradictory. First we use the Ergodicity Breaking parameter $EB(\Delta,t_{exp})$ \cite{He2008,Burov2010,Metzler2017b} which quantifies the dispersion of the time-averaged MSD $\bar{\delta^2}(\Delta,t_{exp})$ \cite{Grebenkov2011b,Grebenkov2011,Sikora2017} with
\begin{equation}
\bar{\delta^2}(\Delta,t_{exp})=\frac{1}{t_{exp}-\Delta}\sum\limits_{n=1}^{t_{exp}-\Delta}(x_{n+\Delta}-x_n)^2
\end{equation} 
 as a function of the experiment duration $t_{exp}$ (i.e. the trajectory length) evaluated with a time-lag $\Delta$:
\begin{equation}
EB(\Delta,t_{exp})=\frac{\langle(\bar{\delta^2}(\Delta,t_{exp}))^2\rangle}{\langle\bar{\delta^2}(\Delta,t_{exp})\rangle^2}-1.
\end{equation}
For an ergodic process, $\lim\limits_{t_{exp} \rightarrow \infty} EB(\Delta,t_{exp})=0$ for any $\Delta$, meaning that for a fixed $\Delta$, the distribution of TAMSD converges to a Dirac delta distribution with $\bar{\delta^2}(\Delta,t_{exp} \rightarrow \infty)=\langle X^2(\Delta)\rangle$.

\begin{figure*}[h!]
\begin{center}  
\includegraphics[width=140mm]{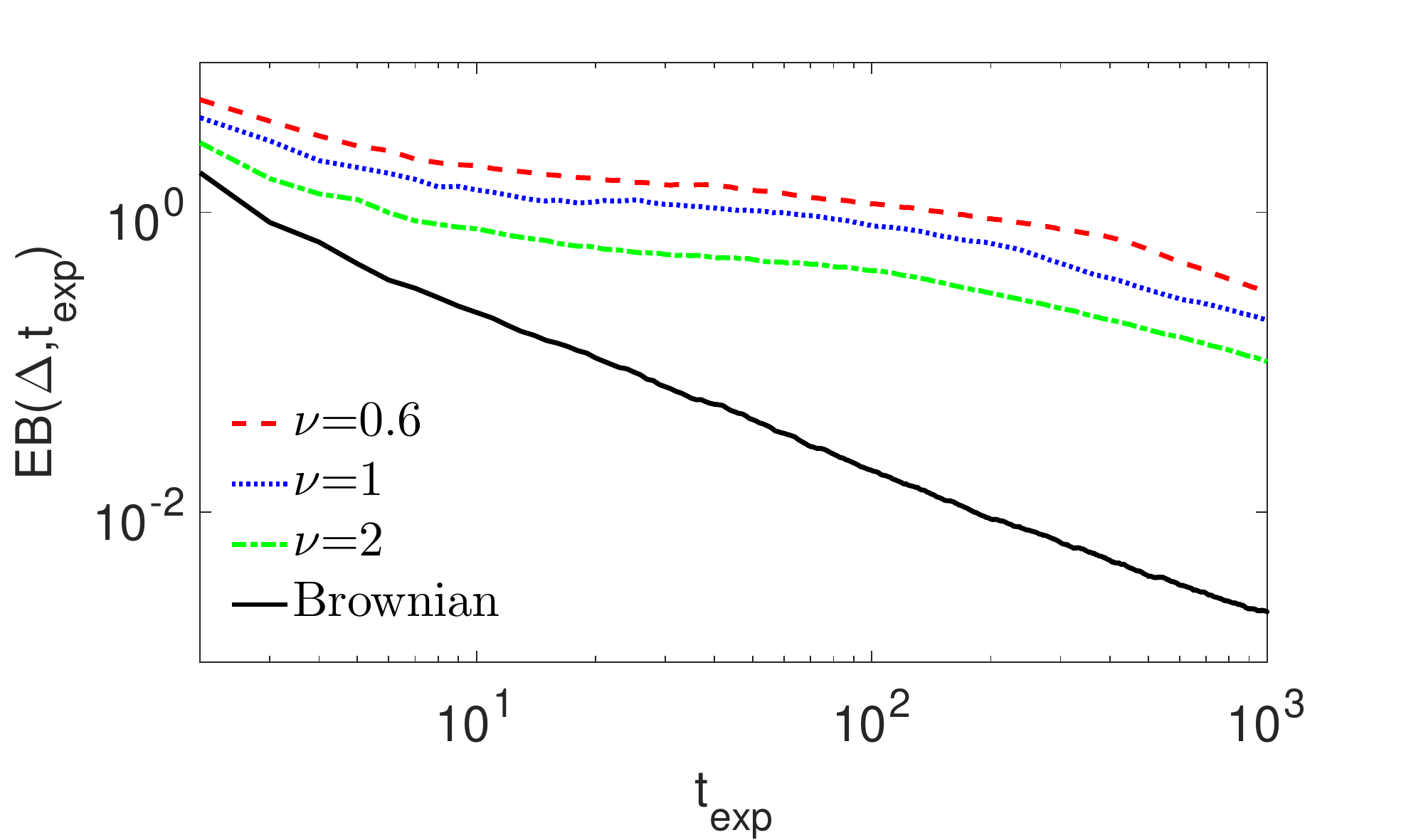}
\\
\includegraphics[width=140mm]{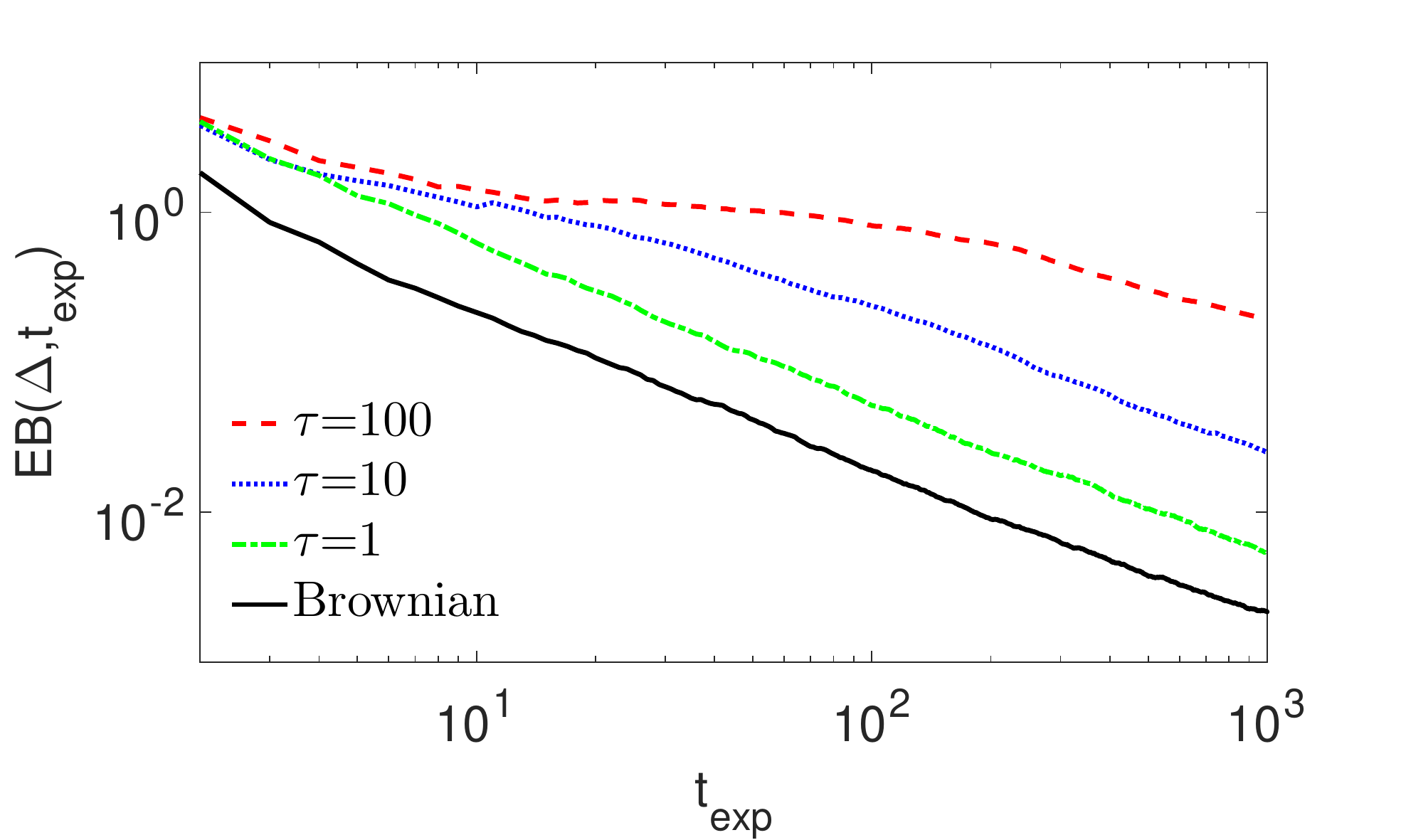} \end{center}
\caption{ Ergodicity breaking parameter calculated by averaging over $M=10^3$ simulated trajectory of length $t_{exp}=10^3$. The result for Brownian motion (solid line) is also plotted for comparison. {\it Top.} Results for variable shape parameter $\nu=\lbrace 0.6,1,2\rbrace$ by varying $\tau$, with $\bar{D}=1$ and $\sigma=1$ being constant.{\it Bottom.}  Results for variable correlation time $\tau=\lbrace 1,10,100\rbrace$ while keeping $\nu=1$, $\bar{D}=1$ and $\sigma=1/\sqrt{\tau}$.}
\label{fig:Ergodicity_Breaking_Parameter}
\end{figure*}

Figure \ref{fig:Ergodicity_Breaking_Parameter} shows that fluctuations of TAMSD are impacted by two characteristics: the shape parameter $\nu$ and the correlation time $\tau$. The smaller the parameter $\nu$, the longer it takes for the EB parameter to vanish. Indeed, for $\nu\leq 1$, diffusivity can be small with high probability that would slow down the dynamics. The correlation time $\tau$ also influences the convergence of $EB(\Delta,t_{exp})$: larger $\tau$ implies longer time to recover from small diffusivities and thus slower dynamics. Setting $\nu=1$ and varying $\tau$, the EB parameter has a transient behavior until $\approx 2\tau$ and decays as a power law $1/t_{exp}$ as in the Brownian case for which the exact formula, in the discrete case, is $EB(\Delta,t_{exp})=\frac{(2\Delta+1/\Delta)}{3(t_{exp}-\Delta+1)}$ \cite{Qian1991}.
Note that a slow decrease of the EB parameter due to disorder was also discussed for fluctuating diffusivity \cite{Cherstvy2016} and diffusion in a periodic potential \cite{Kindermann2016}.

\begin{figure*}[h!]
\begin{center}  
\includegraphics[width=140mm]{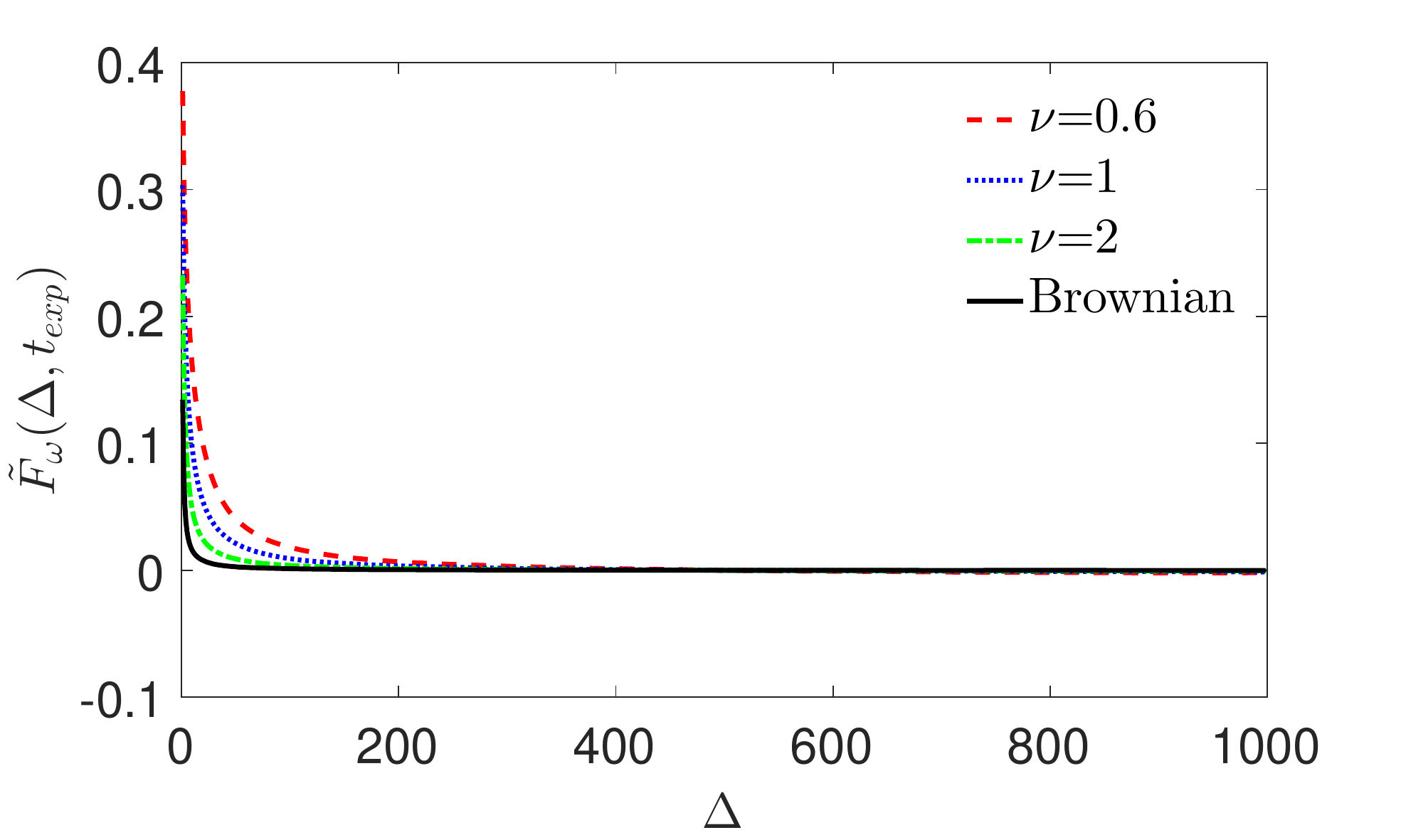}
\\
\includegraphics[width=140mm]{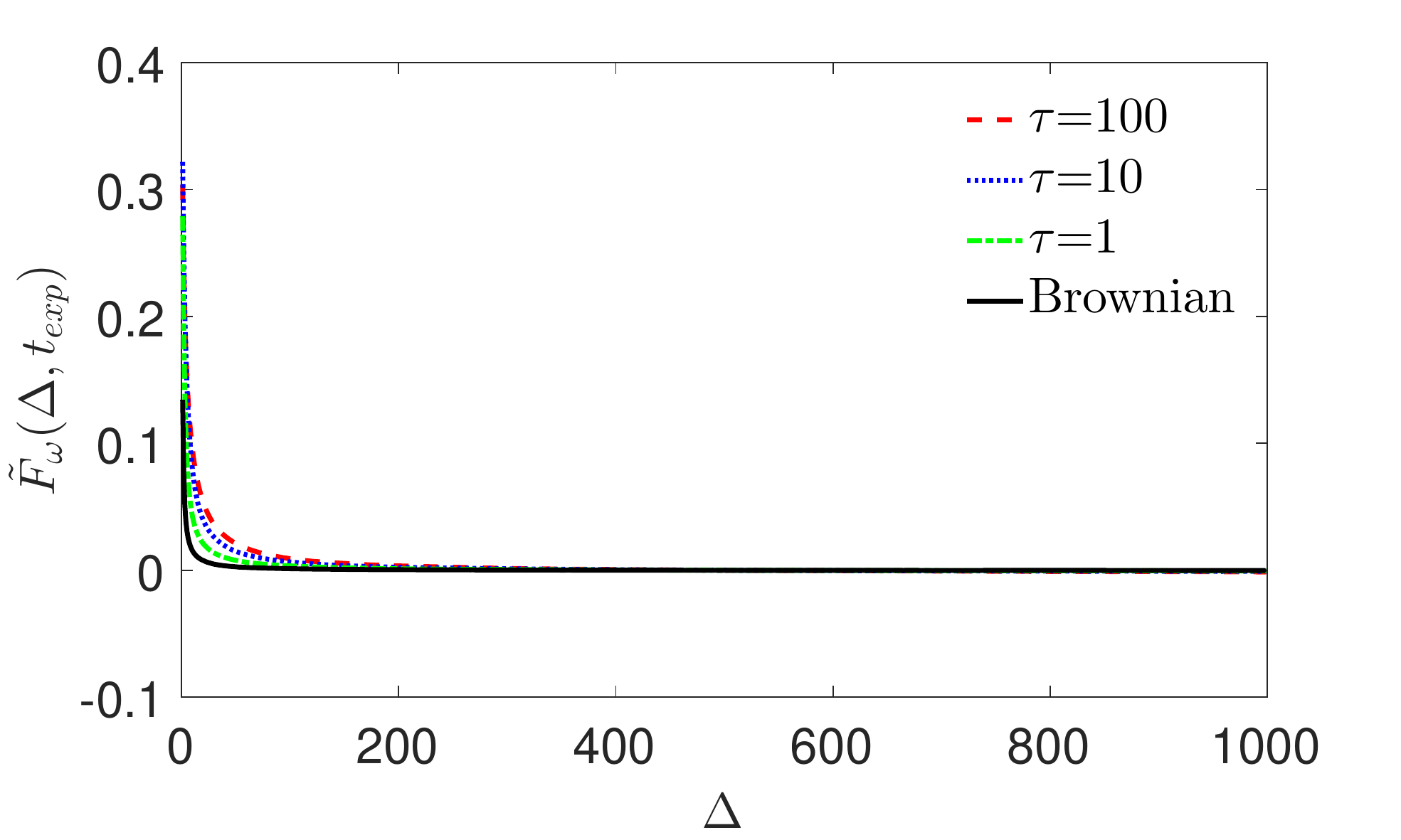}
\end{center}
\caption{ Mean ergodicity estimator $\tilde{F}_\omega(\Delta,t_{exp})$, calculated with $M=10^3$ simulated trajectories of length $t_{exp}=10^3$.  The mean estimator for Brownian motion (solid line) is also plotted for comparison.  {\it Top.} Different values of the shape parameter $\nu=\lbrace 0.6,1,2\rbrace$ by varying $\tau$, with $\bar{D}=1$ and $\sigma=1/\sqrt{\tau}$. {\it Bottom.} Different values of $\tau=\lbrace 1,10,100\rbrace$ while keeping $\nu=1$, $\bar{D}=1$ and $\sigma=1/\sqrt{\tau}$.}
\label{fig:Dynamical_Functional_Parameter}
\end{figure*}

We also discuss the ergodicity test based on the dynamical functional \cite{Magdziarz2011,Lanoiselee2016}. An ergodic process has a vanishing velocity autocorrelation function so that two fragments of the trajectory become independent when time between them is sufficiently long.  The ergodicity estimator $\tilde{F}_\omega(\Delta,t_{exp})$ measures how long it takes before independence is verified on the characteristic function. It has been shown \cite{Magdziarz2011} that for any stationary infinitely divisible ergodic process this function asymptotically vanishes, as also verified by calculating the mean estimator  $\langle\tilde{F}_\omega(\Delta,t_{exp})\rangle$ \cite{Lanoiselee2016} in the case of fractional Brownian motion. In contrast, the mean estimator never vanishes in the case of nonergodic continuous time random walk. In Fig. \ref{fig:Dynamical_Functional_Parameter}, the estimator $\tilde{F}_\omega(\Delta,t_{exp})$ decays fast so that the temporal disorder due to diffusivity does not affect much this quantity, in contrast to the $EB$ parameter. If this estimator vanishes for a single particle trajectory, one can expect asymptotic independence and ergodicity. This implies that getting longer data indeed increases the accuracy of time averaged quantities (smaller $EB(\Delta,t_{exp})$). 
 
The ergodicity breaking parameter shows that the distribution of TAMSD slowly converges to a delta distribution. In turn, the ergodicity estimator $\tilde{F}_\omega(\Delta,t_{exp})$ indicates that the process looses its memory and implies that the TAMSD distribution narrows with increasing $t_{exp}$ (without specifying how). These two quantities do not answer the ergodicity question in the same way, they are complementary. If one needs to know the degree of dispersion of TAMSD, the $EB$ parameter has to be used. The estimator $\tilde{F}_\omega(\Delta,t_{exp})$, which can be applied to a single trajectory, does not quantifies fluctuations, but allows to verify ergodicity, even in the presence of dynamic disorder because it relies on the estimation of the characteristic function of the process.

\section{Discussion}
\label{sec:Discussion}
\subsection{Fourth moment is not enough}\label{subsec:karger_comp}
The K\"{a}rger model \cite{Karger1985} has been developed to study diffusion in a medium in which a particle can randomly switch between two domains with distinct diffusion coefficients $D_1$ and $D_2$, with the exchange rates $K_{12}$ and $K_{21}$.
By solving two coupled diffusion-reaction equations, the Fourier transform of the propagator can be derived \cite{Karger1985}
\begin{equation}
\tilde{P}_{KM}(q,t)=(1-p')\exp(-q^2D_1'(q)t)+p'\exp(-q^2D_2'(q)t),
\end{equation}
with
\begin{eqnarray}
D_1'(q) &=& \frac{1}{2}\left(D_1+D_2+\frac{1}{q^2}(K_{12}+K_{21})\right.\\\nonumber 
&&-\left.
\left(\left(D_2-D_1+\frac{1}{q^2}(K_{21}-K_{12})\right)^2+\frac{4K_{12}K_{21}}{q^4}\right)^{1/2}
\right),
\\\nonumber
D_2'(q)&=& \frac{1}{2}\left(D_1+D_2+\frac{1}{q^2}(K_{12}+K_{21})\right.\\\nonumber
&&+\left.
\left(\left(D_2-D_1+\frac{1}{q^2}(K_{21}-K_{12})\right)^2+\frac{4K_{12}K_{21}}{q^4}\right)^{1/2}
\right),
\\\nonumber
p'&=& \frac{1}{D_2'(q)-D_1'(q)}(p_1D_1+p_2D_2-D_1'(q)),
\end{eqnarray}
where $p_1$ and $p_2$ are relative volume fractions of two domains.
The analytical expression of the non-Gaussian parameter, which was derived in \cite{Jensen2005}, and also studied in \cite{Fieremans2010}, has the same functional form as $\gamma(t)$ from Eq. (\ref{eq:NG_Param}):
\begin{equation}
\gamma_{KM}(t)=\eta \frac{2}{t/\tau}\left(1- \frac{1}{t/\tau}\left( 1-e^{-t/\tau}\right)\right),
\end{equation}
with the coefficient $\eta=\frac{p_1p_2(D_1-D_2)^2}{(p_1D_1+p_2D_2)^2}$,
which corresponds in our case to $\frac{\sigma^2\tau}{\bar{D}}$, and $\tau$ is the exchange time: $\tau=1/K_{12}=1/K_{21}$.

\begin{figure*}[h!]
\begin{center}  
\includegraphics[width=70mm]{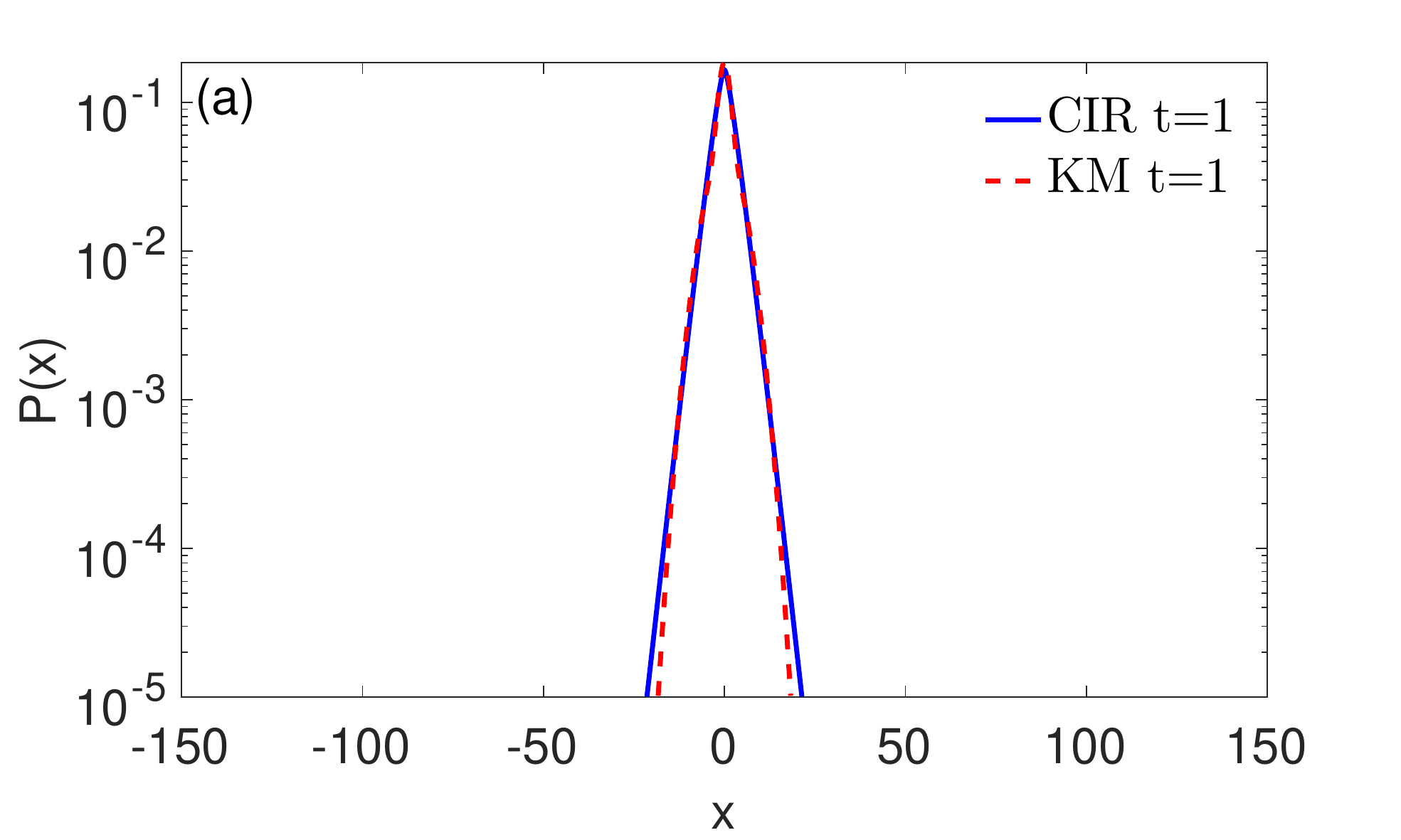}
\includegraphics[width=70mm]{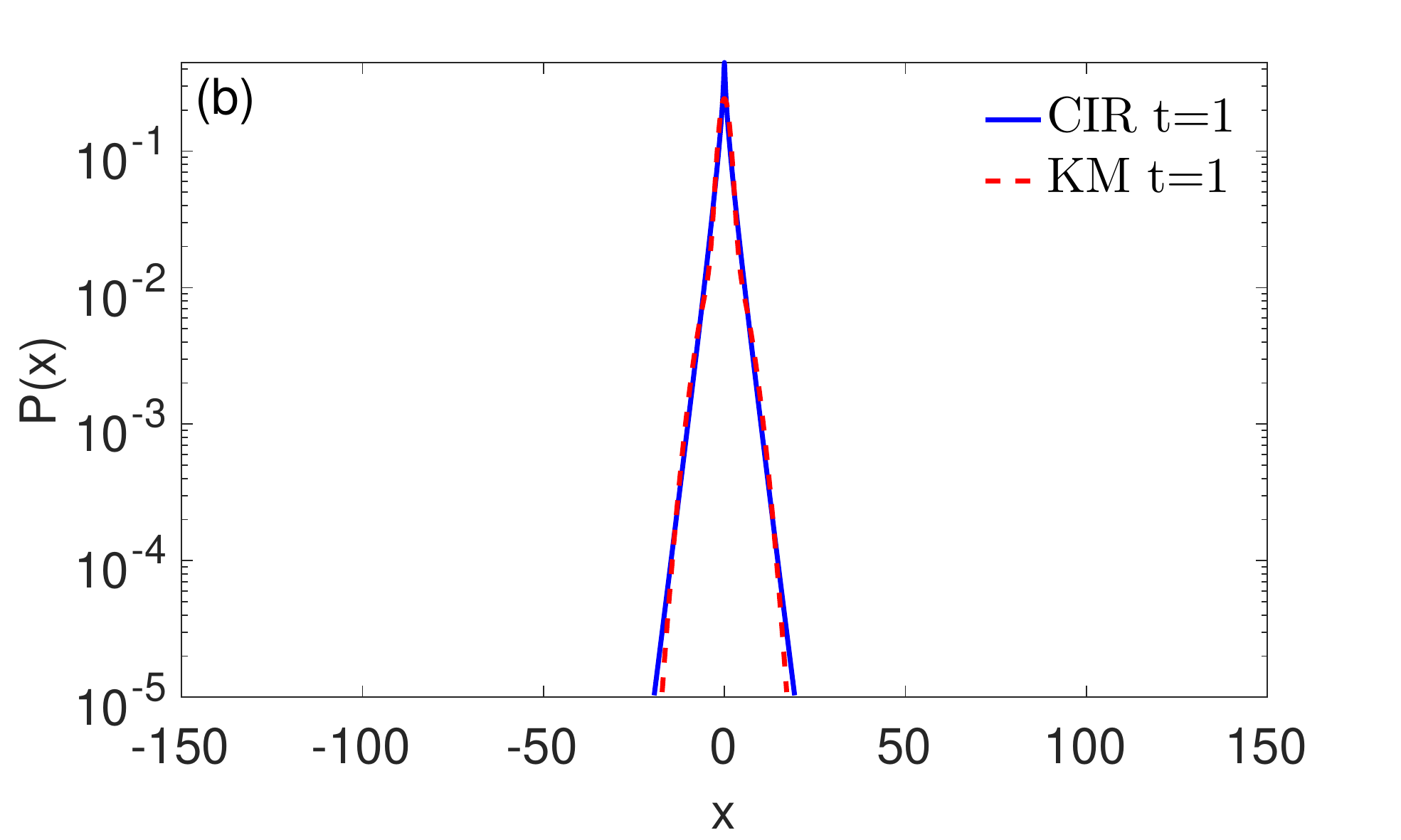}
\includegraphics[width=70mm]{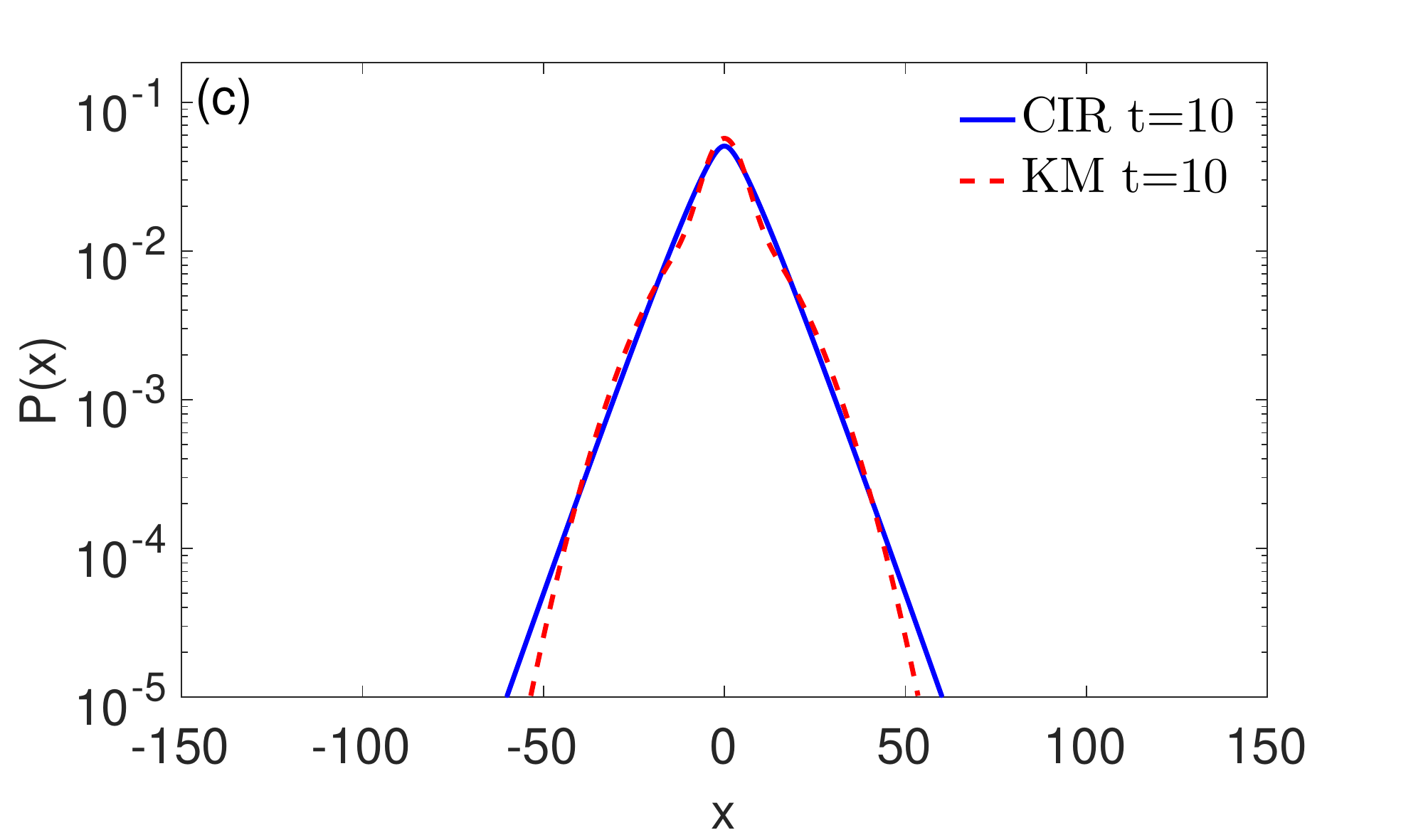}
\includegraphics[width=70mm]{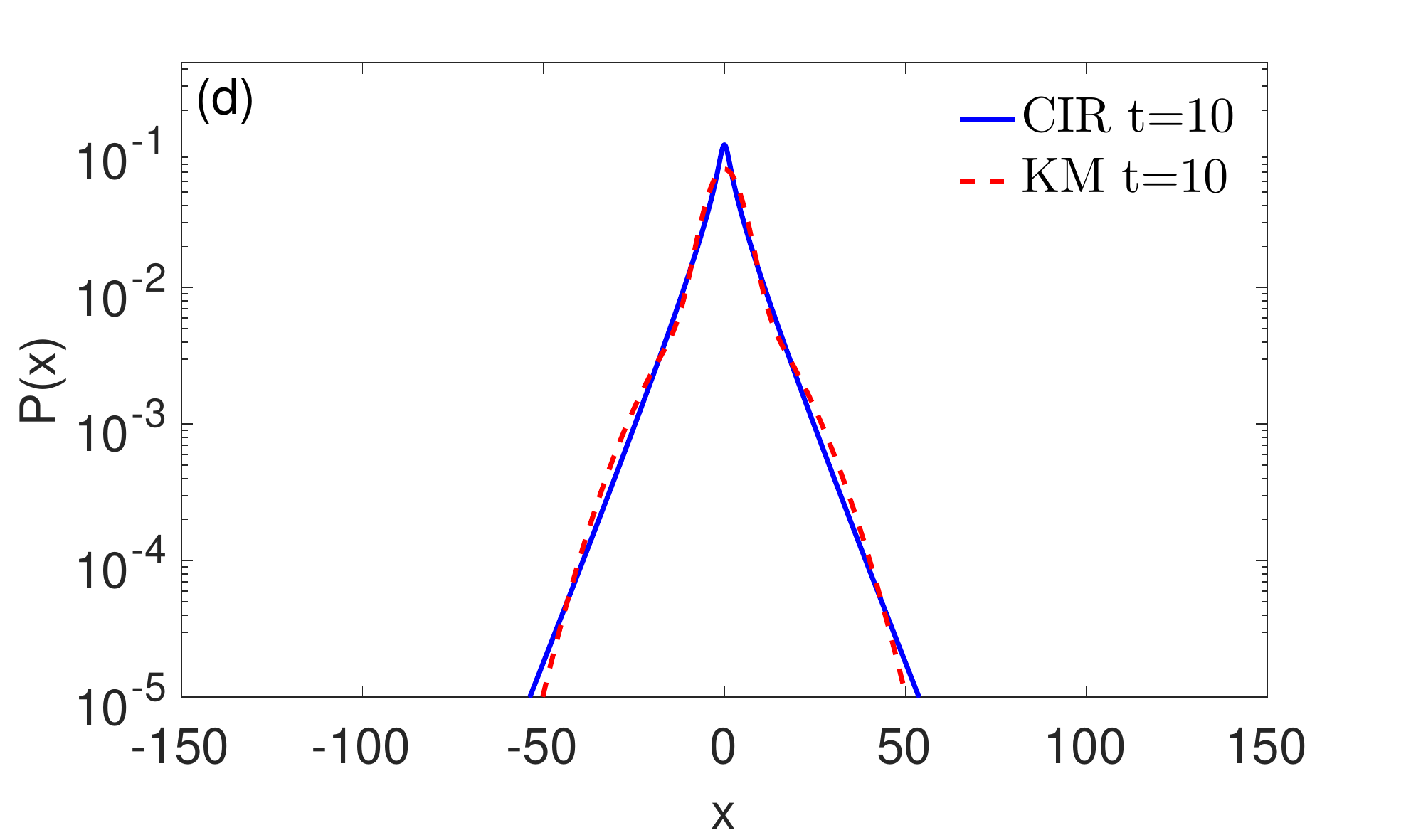}
\includegraphics[width=70mm]{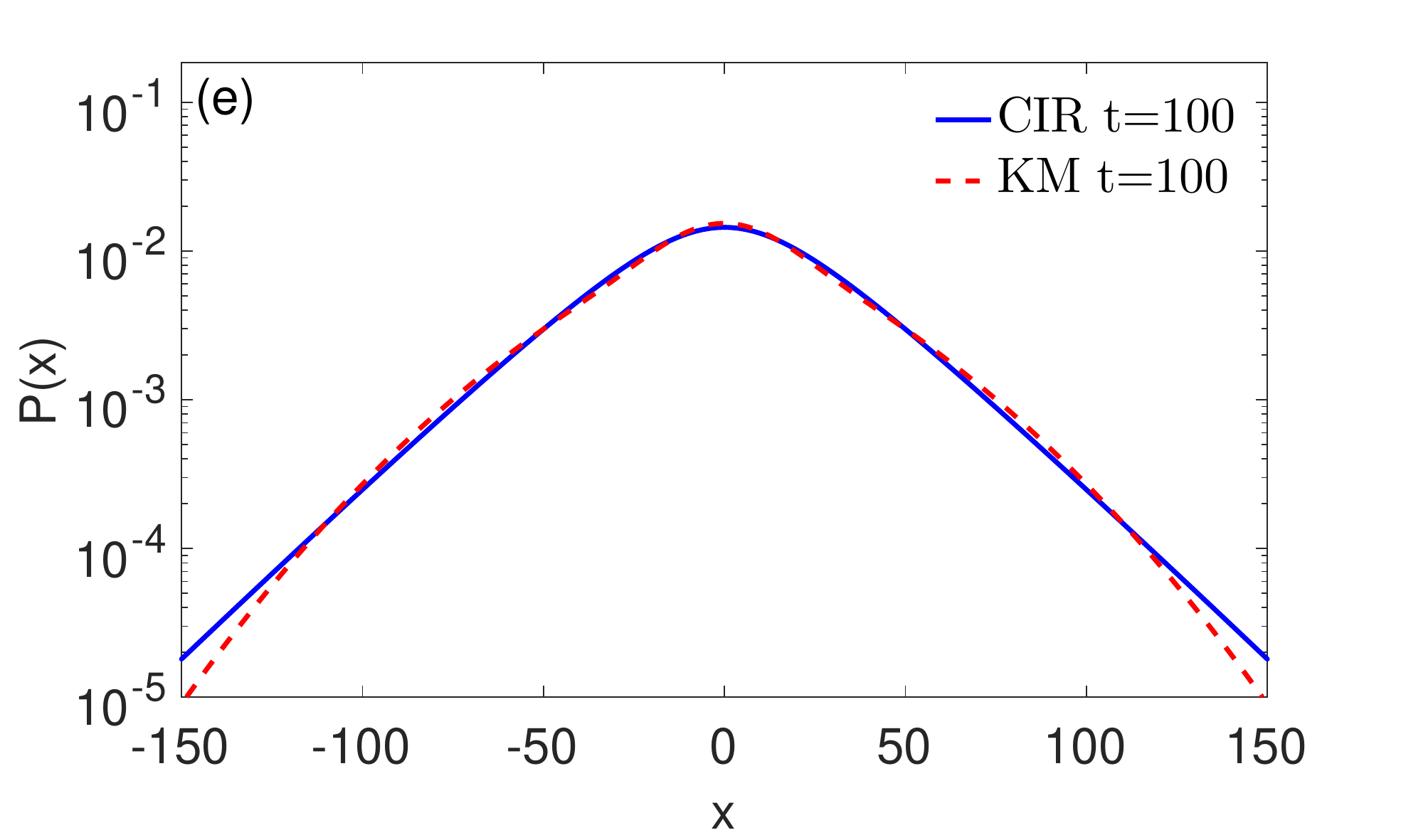}
\includegraphics[width=70mm]{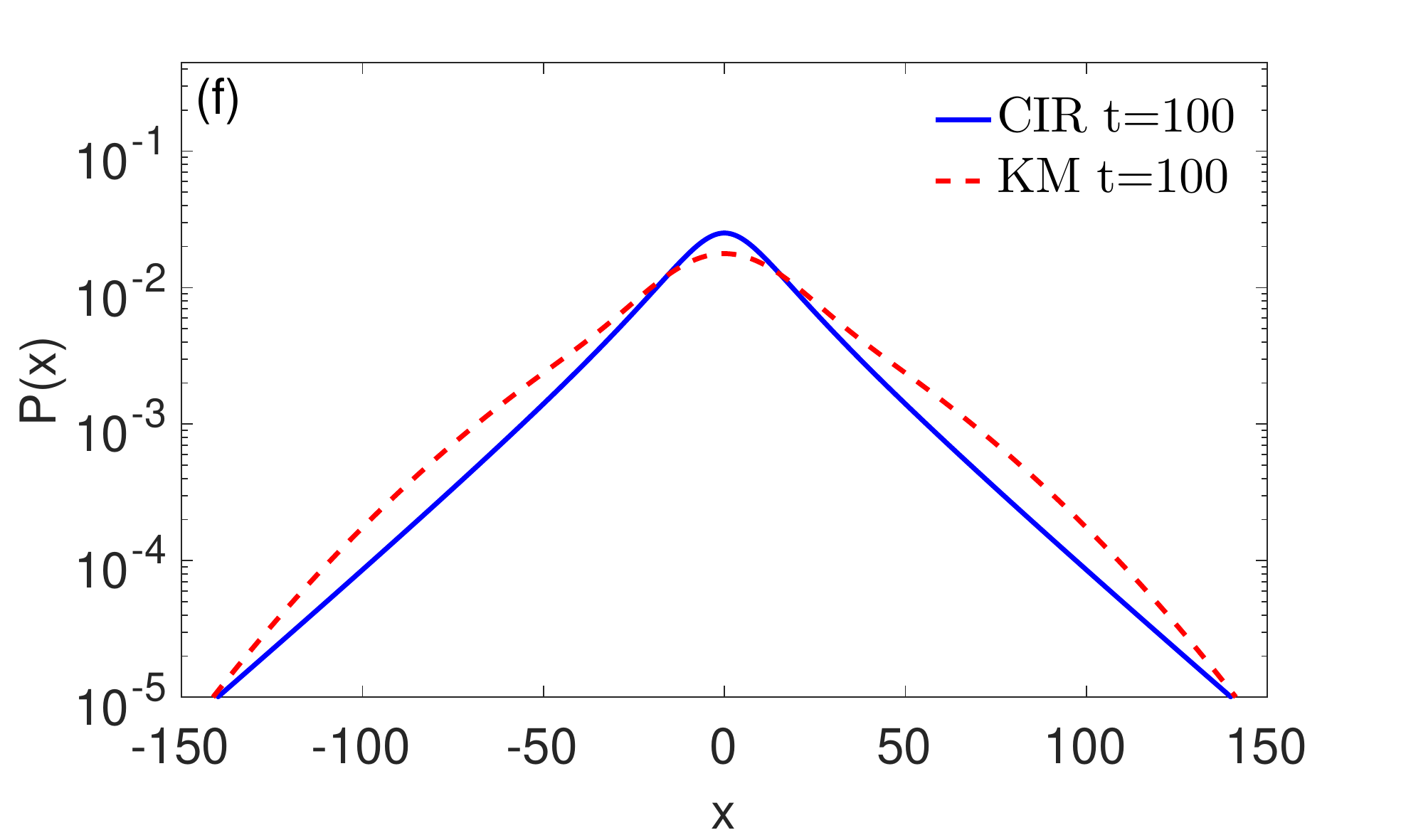}
\end{center}
\caption{Distribution of displacements for the K\"{a}rger model (dashed lines) and our model (solid lines) at time $t=1$ (top), $t=10$ (middle), $t=100$ (bottom). We choose the parameters for the K\"{a}rger model and deduce $\bar{D}=p_1D_1+p_2D_2$ and $\sigma=\sqrt{\bar{D}\eta/\tau}$. {\it Left column.} Parameters of the K\"{a}rger model are $p_1=1/2$, $p_2=1/2$, $D_1=1$ and $D_2=10$ and $\tau=10$, leading to $\nu\approx 1.5$. {\it Right column.} Parameters of the K\"{a}rger model are $p_1=4/5$, $p_2=1/5$, $D_1=1$ and $D_2=10$ and $\tau=10$, leading to $\nu\approx 0.6$.}
\label{fig:Comparaion_KM_CIR}
\end{figure*}

Figure \ref{fig:Comparaion_KM_CIR} compares distributions for the K\"{a}rger model and our approach. In the case $\nu>1$, both distributions are very close at all times. In the case $\nu\leq 1$, obtained here by setting different relative volumes $p_1$ and $p_2$, the K\"{a}rger model does not reproduce the peak at $0$. In other words, the K\"{a}rger model as a superposition of only two Gaussian distributions does not match our model with infinitely many Gaussian distributions (see Sec. \ref{subsec:Superstatistics_Short_Time}).

We conclude that these two distributions having identical second and fourth moments, are still different. While it was known that mean squared displacement is not sufficient to characterize a model, here we illustrate that even the fourth moment (and the non-Gaussian parameter) is not enough. 


\subsection{Anomalous diffusion}
In biology there are many experimental evidences of anomalous diffusion \cite{Weiss2004,Barkai2012,Hofling2013,Manzo2015,Sadegh2017}  when the mean squared displacement scales as a power law with time $\langle X^2(t)\rangle=2D_\alpha t^\alpha$, where $D_\alpha$ is the generalized diffusion coefficient and $\alpha$ is the anomalous exponent. 
We propose an extension by a simple scaling of the time $t/\tau\Rightarrow (t/\tau)^\alpha$, so that $\tilde{P}(q,t)$ from Eq. (\ref{eq:margin_sol_no_init}) is replaced by
\begin{eqnarray}
\fl
\tilde{P}(q,t)&=&
\left(e^{-\frac{1}{2}\left(\omega-1\right) (t/\tau)^\alpha}
\frac{4\omega}{\left(\omega+1\right)^2}\left(1-\left(\frac{\omega-1}{\omega+1}\right)^2
e^{-\omega (t/\tau)^\alpha} \right)^{-1}
\right)^\nu.
\end{eqnarray} 
The non-Gaussian parameter would now depend on $\alpha$:
\begin{equation}
\gamma(t)=\frac{ 2 \sigma^2\tau^{1+\alpha}}{ \bar{D} t^\alpha} \left(1-\frac{1}{(t/\tau)^\alpha}\left( 1-e^{-(t/\tau)^\alpha}\right)\right).
\end{equation}
As expected, in the subdiffusive case $\alpha<1$, the convergence to a Gaussian distribution is slower as compared to the superdiffusive case $\alpha>1$, because larger $\alpha$ means faster exploration of space. 

\section{Conclusion}
We presented a model of non-Gaussian diffusion, based on coupled Langevin equations. We derived the explicit exact formula for the distribution of displacements in the Fourier-Laplace domain and studied different asymptotic regimes. We showed that this distribution exhibits exponential tails and converges slowly, as $1/t$, to a Gaussian one. Depending on the shape parameter $\nu$, the distribution can be flat ($\nu>1$) or peaked ($\nu\leq 1$) at zero. 
The MSD evolves linearly with time in spite of non-Brownian character of the motion. We pointed that the ergodicity estimator $\tilde{F}_\omega(\Delta,t_{exp})$ catches ergodic nature of the process while the random nature of diffusivity makes fluctuations of TAMSD to span up at long times as demonstrated by the ergodicity breaking parameter $EB(\Delta,t_{exp})$ making of the TAMSD a bad estimator of the average diffusion coeffficient $\bar{D}$. We used the autocorrelation of squared increments to determine the autocovariance structure of diffusivity. Given that small diffusivities are made much more probable in the case $0<\nu<1$ (which was not accessible in former models), the underlying process exhibits a richer phenomenology. We expect that this model will help to understand more deeply dynamical heterogeneities observed in experiments. An important perspective is to relate the correlation structure of the stochastic diffusivity $D_t$ with spatial correlations structure of the medium \cite{Valle1991}. One can also analyze the first passage time (FPT) statistics in our model of heterogeneous diffusion to reveal the impact of the diffusing diffusivity. Although the mean squared displacement grows linearly with time, the distribution of FPT can be sensitive to the related annealed disorder (e.g., see \cite{Hernandez1990a,Hernandez1990b} for models of quenched disorder).
\section*{Acknowledgments}
DG acknowledges the financial support by French National Research
Agency (ANR Project ANR-13-JSV5-0006-01).
The authors thank Aleksei Chechkin, Marcel Filoche, Nicolas Moutal, Dimitri Novikov and David Waxman for inspiring discussions.
\appendix
\section{Derivation of the Cox-Ingersoll-Ross equation}
\label{sec:BDeriv_CIR}

Let us consider a collection of independent Ornstein-Uhlenbeck processes indexed by $i\in\left[1,n\right]$, each of them obeying the Langevin equation:
\begin{equation}
dY_t^{(i)}=-\frac{1}{2}\beta Y_t^{(i)}dt+\sigma dW_t^{(i)},
\end{equation}
where $\beta$ is the inverse correlation time, $\sigma$ is the level of fluctuations, and $W_t^{(i)}$ are independent Wiener processes.
Following \cite{Jain2016,Chechkin2016} the diffusing diffusivity is modeled as  
\begin{equation}
D_t=\sum\limits_{i=1}^n\left(Y_t^{(i)}\right)^2.
\end{equation}
Let $f(y_1,y_2...,y_n)=\sum\limits_{i=1}^n y_i^2$ so that $\frac{\partial}{\partial y_i}f=2y_i$ and $\frac{\partial^2}{\partial y_i\partial y_j}f=2\delta_{ij}$.

According to It\^o formula, we get
\begin{eqnarray}\nonumber
\fl
dD_t=\sum\limits_{i=1}^n \frac{\partial}{\partial y_i}fdY_t^{(i)}+\frac{1}{2}\sum\limits_{i,j=1}^n\frac{\partial^2}{\partial y_i\partial y_j}fdY_t^{(i)}dY_t^{(j)}\\\nonumber
\fl
\quad\quad=\sum\limits_{i=1}^n 2Y_t^{(i)}\left(-\frac{1}{2}\beta Y_t^{(i)}dt+\sigma dW_t^{(i)}\right)+\sum\limits_{i=1}^n\left(-\frac{1}{2}\beta Y_t^{(i)}dt+\sigma dW_t^{(i)}\right)^2\\\nonumber
\fl
\quad\quad=\sum\limits_{i=1}^n 2Y_t^{(i)}\left(-\frac{1}{2}\beta Y_t^{(i)}dt+\sigma dW_t^{(i)}\right)+n\sigma^2 dt\\\nonumber
\fl
\quad\quad=\left(n\sigma^2-\beta\sum\limits_{i=1}^n  \left(Y_t^{(i)}\right)^2\right)dt+2\sigma\sum\limits_{i=1}^n Y_t^{(i)}dW_t^{(i)}\\\nonumber
\fl
\quad\quad=\left(n\sigma^2 -\beta D_t\right)dt+2\sigma\sum\limits_{i=1}^n Y_t^{(i)}dW_t^{(i)}\\\nonumber
\fl
\quad\quad=\left(n\sigma^2 -\beta D_t\right)dt+2\sigma\sqrt{D_t}\sum\limits_{i=1}^n \frac{Y_t^{(i)}}{\sqrt{D_t}}dW_t^{(i)}.\\\nonumber
\end{eqnarray}

The stochastic process $W_t$ defined as $W_t=\sum\limits_{i=1}^n \int_0^t\frac{Y_s^{(i)}}{\sqrt{D_s}}dW_s^{(i)}$ is a martingale because it has no drift \cite{McCauley2013}. For its increments, $dW_t=\sum\limits_{i=1}^n \frac{Y_t^{(i)}}{\sqrt{D_t}}dW_t^{(i)}$, we verify that $dW_t dW_t=\sum\limits_{i=1}^n \frac{\left(Y_t^{(i)}\right)^2}{D_t}\left(dW_t^{(i)}\right)^2=dt$, so the increments are properly normalized. We conclude that $W_t$ is a Wiener process.

We now can rewrite the above equation as:
\begin{equation}
dD_t=\left(n\sigma^2 -\beta D_t\right)dt+\sqrt{2}\sigma\sqrt{2D_t}dW_t.
\end{equation}
Setting $\tilde{\sigma}=\sqrt{2}\sigma$, $\beta=1/\tau$ and $n=\frac{\bar{D}}{\tilde{\sigma}^2\tau}=\nu$
one finally retrieves the Cox-Ingersoll-Ross equation
\begin{equation}
dD_t=\frac{1}{\tau}\left(\bar{D} - D_t\right)dt+\tilde{\sigma}\sqrt{2D_t}dW_t.
\end{equation}

\section{Solution of the Fokker-Planck equation}
\label{sec:ASolution_Equation}
\subsection{Statement of the problem}
 We consider the two-dimensional forward Fokker-Planck equation (\ref{FP_xD}) on position and diffusivity with the initial condition $P(x,D,t=0\vert x_0,D_0)=\delta(x-x_0)\delta(D-D_0)$.\\
The equation can be formally expressed in term of a two-dimensional probability density flux $\vec{J}=\lbrace J_x,J_D \rbrace$
\begin{equation}\label{eq:Fokker_Equa_flux}
\frac{\partial }{\partial t}P(x,D,t\vert x_0,D_0)=-\textrm{div}\left(\vec{J}\right),
\end{equation}
 with components $J_x=-D\frac{\partial }{\partial x}P$ and $J_D=-\frac{1}{\tau}\left(D-\bar{D}\right)P-\sigma^2\frac{\partial }{\partial D}\left[DP\right]$.
This equation can be solved by transforming the position $x$ into the Fourier space, and the diffusivity, defined on the real half line $D\in [0,\infty)$, into the Laplace space:
\begin{equation}
\fl
\tilde{P}(q,s,t\vert x_0,D_0)=\int_{-\infty}^\infty dxe^{-iqx}\int_0^\infty dDe^{-Ds} P(x,D,t\vert x_0, D_0).
\end{equation}
so that Eq. (\ref{eq:Fokker_Equa_flux}) becomes:
\begin{equation}
\frac{\partial }{\partial t}\tilde{P}=\int_{-\infty}^\infty dxe^{-iqx}\int_0^\infty dDe^{-Ds}\left(D\frac{\partial^2 }{\partial x^2}P-\frac{\partial }{\partial D}J_D\right)
\end{equation}
This leads to the first-order partial differential equation:
\begin{equation}\label{eq:ann_TF_TL_flux}
\frac{\partial}{\partial t}\tilde{P}+G(s)\frac{\partial}{\partial s}\tilde{P}=-\frac{\bar{D}}{\tau} s\tilde{P}
+J_D(D=0,t),
\end{equation}
with $G(s)=\sigma^2\left(s-s_1\right)\left(s-s_2\right)$, where $s_1=\frac{-1+\omega}{2\sigma^2\tau}$, $s_2=\frac{-1-\omega}{2\sigma^2\tau}$, and
$\omega=\sqrt{1+4\sigma^2\tau^2q^2}$.
 The initial condition is now $\tilde{P}(q,s,t=0\vert x_0, D_0)=e^{-iqx_0}e^{-sD_0}$.
 The last term in Eq. (\ref{eq:ann_TF_TL_flux}) is the probability density flux at $D=0$ which can be equivalently written $J_D(D=0,t)=\sigma^2(\nu-1)P(q,D=0,t\vert x_0,D_0)$.
 First we solve the problem for the model of reflecting boundary in \ref{sec:sol_reflect_bound} and then demonstrate the effect with absorbing boundary condition at $D=0$ in \ref{sec:solution_absorbing_boundary}.
\subsection{Reflecting boundary condition at $D=0$}
\label{sec:sol_reflect_bound}
First we solve the problem in the case of reflecting boundary condition, i.e. without flux at $D=0$: $J_D(D=0,t)=0$. For $\nu\geq 0$, we search for the solution of the equation
\begin{equation}
\frac{\partial}{\partial t}\tilde{P}+G(s)\frac{\partial}{\partial s}\tilde{P}=-\frac{\bar{D}}{\tau} s\tilde{P},
\end{equation}
in the form:
\begin{equation}
\label{form_sol}
\tilde{P}(q,s,t\vert x_0,D_0)=f\left(t-g\left(s\right)\right)h\left(s\right),
\end{equation}
with three unknown functions $f,g,h$.
Nontrivial solutions are found by solving 
\begin{equation}
\left\{
    \begin{array}{ll}
1-g'G=0, \\\\
Gh'+\frac{\bar{D}}{\tau} sh=0,
    \end{array}
\right.
\end{equation}
which gives
\begin{equation}
\left\{
    \begin{array}{ll}
g(s)=\frac{\tau}{\omega}\ln \left(\frac{s-s_1}{s-s_2}\right),\\\\
h(s)=(s-s_1)^{\frac{-\bar{D}}{\omega}s_1}(s-s_2)^{\frac{\bar{D}}{\omega}s_2}.
    \end{array}
\right.
\end{equation}
Now we use the initial condition to deduce the function $f$:
\begin{equation}
\tilde{P}(q,s,t=0\vert x_0,D_0)=e^{-iqx_0}e^{-sD_0}=f\left(-g\left(s\right)\right)h\left(s\right),
\end{equation}
from which we get
\begin{equation}
f(z)=\frac{e^{-iqx_0}e^{-D_0g^{-1}(-z)}}{h\left(g^{-1}\left(-z\right)\right)},
\end{equation}
or equivalently,
\begin{equation}
f(z)=\left(\frac{\omega}{\sigma^2\tau}\right)^{\nu}e^{-iqx_0}\exp\left(-D_0\frac{s_1-s_2e^{-\omega z/\tau}} {1-e^{-\omega z/\tau}}\right)(1-e^{-\omega z/\tau})^{-\nu}e^{-\bar{D}s_1 z/\tau }.
\end{equation}
The solution is finally
\begin{eqnarray}
\fl\nonumber
\tilde{P}(q,s,t\vert x_0, D_0)&=&F(x_0,D_0,s)\left(\frac{\omega}{\sigma^2\tau}\right)^{\nu}
\left(s-s_2-(s-s_1)e^{-\omega t/\tau}\right)^{-\nu}\\\label{eq:sol_s1s2}
&&\times\exp\left(-\bar{D} \left(\frac{\omega-1}{2\sigma^2\tau}\right) t/\tau\right),
\end{eqnarray}
with $\nu$ from Eq. (\ref{eq:nu}) and 
\begin{equation}\label{eq:sol_s1s2_x0D0}
F(x_0,D_0,s)=
\exp\left(-iqx_0-D_0\frac{s_1-s_2\frac{s-s_1}{s-s_2}
e^{-\omega t/\tau}} {1-\frac{s-s_1}{s-s_2}e^{-\omega t/\tau}}\right).
\end{equation}
Substituting $s_1$ and $s_2$ in Eq. (\ref{eq:sol_s1s2}) and Eq. (\ref{eq:sol_s1s2_x0D0}), we get Eq. (\ref{eq:complete_solution}). 

In practice, it is hard to access directly the time-dependent diffusivity. It is therefore convenient to integrate the joint probability density over diffusivity to get the marginal distribution $\tilde{P}(q,t\vert x_0,D_0)$, which can be obtained in the Laplace domain by simply setting $s=0$:

\begin{eqnarray}
\fl\nonumber
\tilde{P}(q,t\vert x_0,D_0)&=&
F(x_0,D_0,s=0 |x_0)
\left(\frac{
\left(\frac{2\omega}{1+\omega}\right)}{1-\left(1-\frac{2\omega}{1+\omega}\right)e^{-\omega t/\tau}
}\right)^{\nu}\\
&&\times\exp\left(-\bar{D} \left(\frac{\omega-1}{2\sigma^2\tau}\right) t/\tau\right).
\end{eqnarray}

Another issue is the dependence on the initial diffusivity $D_0$. If the system is in a stationary regime for the diffusivity, one can average over the stationary distribution $\Pi(D_0)$.
This distribution can be obtained from Eq. (\ref{eq:complete_solution}) by averaging over position (i.e. by setting $q=0$), then taking the limit $t\rightarrow\infty$ and using the inverse Laplace transform relation 
\begin{equation}
\Pi(D_0)=\mathcal{L}^{-1} \left[\left(s+\frac{1}{\sigma^2\tau}\right)^{-\nu}\right] ={\frac{\nu^\nu}{\Gamma(\nu)\bar{D}^\nu}D_0^{\nu-1}\exp\left(-\frac{D_0}{\sigma^2\tau}\right)}, 
\end{equation}
also known from \cite{Feller1951}. Then, the average over initial diffusivity reads
\begin{equation}\label{eq:form_average_D0}
\tilde{P}(q,t\vert x_0)=\int\limits_0^\infty\Pi(D_0)\tilde{P}(q,t\vert x_0,D_0)dD_0.
\end{equation}
Taking the integral, we deduce Eq. (\ref{eq:margin_sol_no_init}).
\subsection{Absorbing boundary condition at $D=0$ }
\label{sec:solution_absorbing_boundary}
\begin{figure*}[h!]
\begin{center}  
\includegraphics[width=140mm]{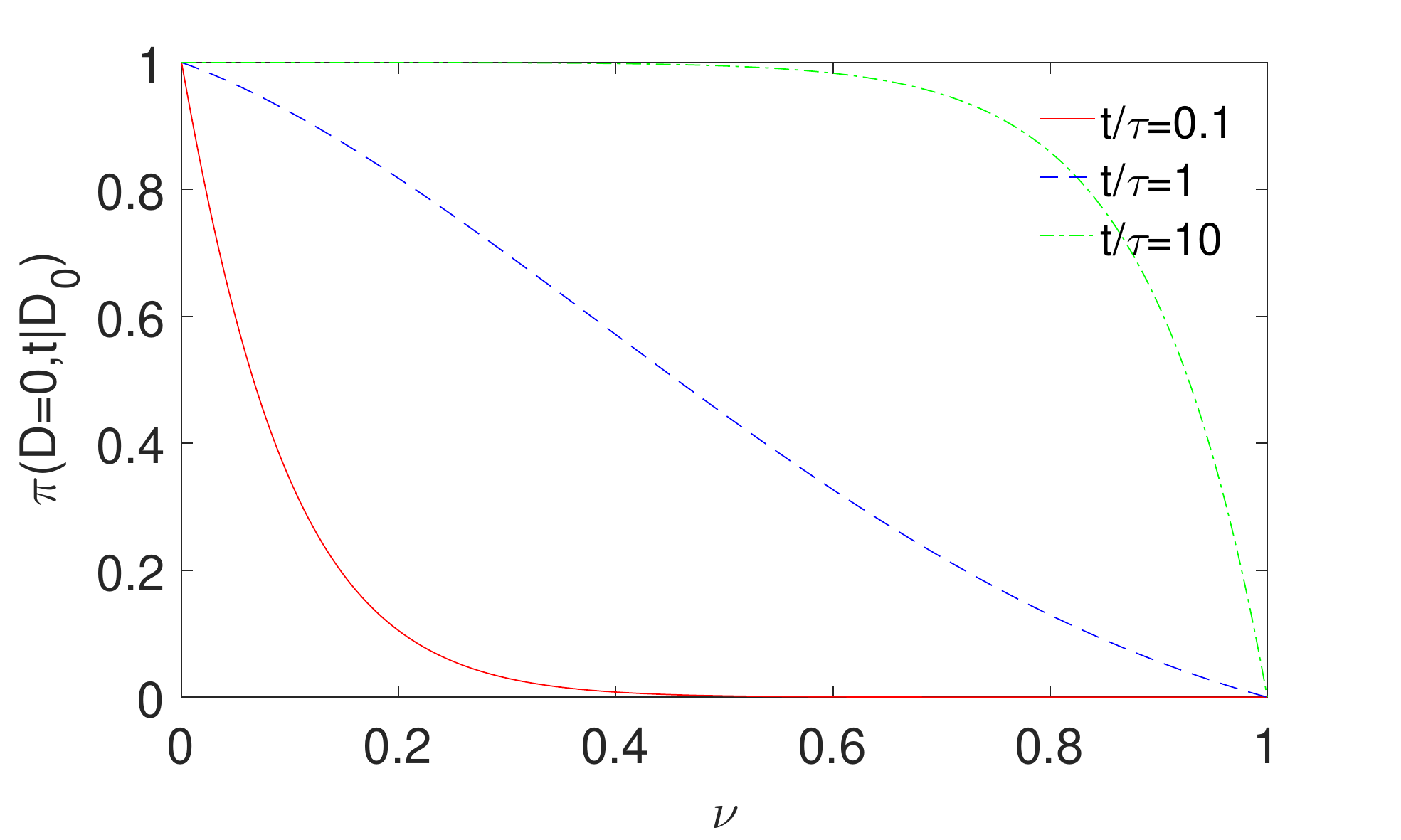}
\end{center}
\caption{Probability of $D=0$ as a function of $\nu$ for three different dimensionless times $t/\tau=\{ 0.1,1,10\}$. $\nu$ is varied from $0$ to $1$ by keeping $\sigma=1$, $\tau=10$ and adjusting $\bar{D}=\nu\sigma^2\tau$ with initial diffusivity $D_0=\bar{D}$.}
\label{fig:prob_D_equal_zero}
\end{figure*}

In the case with absorbing boundary condition at $D=0$, the Fokker-Planck equation represents the evolution with time of the probability density of being at position $x$ with diffusivity $D$ starting at $x_0$, $D_0$, without ever having diffusivity $D=0$. In other words, we look at the propagator of particles which have never stopped. For simplicity we adopt the notation $J_D(D=0,t)=\sigma^2(\nu-1)\phi(t)$.
  This problem is solved using the method of characteristics. The solution of the equation
 \begin{equation}\label{equa_carac}
 dt=\frac{ds}{G(s)}=\frac{d\tilde{P}}{\sigma^2(\nu-1)\phi(t)-\frac{s\bar{D}}{\tau}\tilde{P}}
 \end{equation}
gives
\begin{equation}\label{C1}
s=\frac{s_1-s_2C_1e^{\omega t/\tau}}{1-C_1e^{\omega t/\tau}},
\end{equation}
with a constant $C_1$. This expression is used to deduce the homogeneous solution of the second equation in Eq. (\ref{equa_carac}):
\begin{equation}
\tilde{P}_h(q,s,t)=\left[C_1e^{\omega t/\tau}\right]^{-\frac{\nu}{2}\left(1-\frac{1}{\omega}\right)}
\left[1-C_1e^{\omega t/\tau}\right]^{\nu}.
\end{equation}
Together with the particular solution we find
\begin{eqnarray}\label{Ph_Ppar}
\fl
\tilde{P}(q,s,t\vert x_0,D_0)&=&\tilde{P}_h(q,s,t)\\\nonumber
\fl &&\times
\left[C_2+\sigma^2(\nu-1)\int\limits_0^{t}dt^{'}\frac{\phi(t^{'})}{\left[C_1e^{\omega t^{'}/\tau}\right]^{-\frac{\nu}{2}\left(1-\frac{1}{\omega}\right)}
\left[1-C_1e^{\omega t^{'}/\tau}\right]^{\nu}}\right].
\end{eqnarray}
We pose that $C_2$ is an arbitrary function $H$ of $C_1$: $C_2=H(C_1)$. The initial condition determines the function $H$:
\begin{equation}\label{C2}
H(u)=e^{-iqx_0}\exp\left(-D_0\frac{s_1-s_2u}{1-u}\right)
\left[u\right]^{\frac{\nu}{2}\left(1-\frac{1}{\omega}\right)}
\left[1-u\right]^{-\nu}.
\end{equation}
For brevity we make the substitution $\rho=\frac{s-s_1}{s-s_2}$. Inserting Eq. (\ref{C1}) and Eq. (\ref{C2}) to Eq. (\ref{Ph_Ppar}) leads to the solution
\begin{eqnarray}
\fl
\tilde{P}(q,s,t\vert x_0,D_0)&=& e^{-iqx_0}\exp\left(-D_0\frac{s_1-s_2\rho e^{-\omega t/\tau}}{1-\rho e^{-\omega t/\tau}}\right)\left(e^{-\omega t/\tau}\right)^{\frac{\nu}{2}\left(1-\frac{1}{\omega}\right)}\left(\frac{1-\rho e^{-\omega t/\tau}}{1-\rho}\right)^{-\nu}\\\nonumber
\fl
&&+
\sigma^2(\nu-1)\int\limits_0^{t}dt^{'}\frac{\phi(t^{'})}{\left(e^{-\omega(t-t^{'})/\tau}\right)^{-\frac{\nu}{2}\left(1-\frac{1}{\omega}\right)}
}\left(\frac{1-\rho e^{-\omega(t-t^{'})/\tau}}{1-\rho}\right)^{-\nu},
\end{eqnarray}
with the function $\phi(t)$ to be determined. The approach by Feller \cite{Feller1951} is to require that the inverse Laplace transform of the propagator exists. At large $s$, $\tilde{P}$ becomes
\begin{eqnarray}
\fl
\tilde{P}(q,s\to\infty,t\vert x_0,D_0)&=&
s^{-\nu}
\left(\frac{1-e^{-\omega t/\tau}}{s_1-s_2}\right)^{-\nu}
\left(e^{-\omega t/\tau}\right)^{\frac{\nu}{2}\left(1-\frac{1}{\omega}\right)}\\\nonumber
&&\times\Bigg[
e^{-iqx_0}\exp\left(-D_0\frac{s_1-s_2\rho e^{-\omega t/\tau}}{1-\rho e^{-\omega t/\tau}}\right)\\\nonumber
&&\quad\quad+\sigma^2(\nu-1)\int\limits_0^{t}dt^{'}\frac{\phi(t^{'})}{\left(e^{-\omega t^{'}/\tau}\right)^{\frac{\nu}{2}\left(1-\frac{1}{\omega}\right)}
}\left(\frac{1-e^{-\omega(t-t^{'})/\tau}}{1-e^{-\omega t/\tau}}\right)^{-\nu}
\Bigg].
\end{eqnarray}
At this stage one sees that for $\nu\geq 1$,  the existence of the inverse Laplace transform is ensured by the $s^{-\nu}$ factor so taking $\phi(t)=0$ gives the exact solution, and the particle never gets $D=0$. When $\nu<1$, the finite solution at $D=0$ exists only if the sum cancels in the right hand side, leading to the integral equation (cf. \cite{Feller1951}):
\begin{equation}
\fl
\int\limits_0^{t}dt^{'}\frac{\phi(t^{'})}{\left(e^{\omega y/\tau}\right)^{\frac{\nu}{2}\left(1-\frac{1}{\omega}\right)}}
\left(\frac{1-e^{-\omega(t-t^{'})/\tau}}{1-e^{-\omega t/\tau}}\right)^{-\nu}
=-\frac{e^{-iqx_0}}{\sigma^2(\nu-1)}\exp\left(-D_0\frac{s_1-s_2 e^{-\omega t/\tau}}{1- e^{-\omega t/\tau}}\right).
\end{equation}
Using the variable change $1-e^{-\Omega t}=z^{-1}$ and $1-e^{-\Omega y}=\xi^{-1}$
and defining
\begin{equation}
g(u)=\frac{\tau}{\omega}\phi(u)e^{iqx_0}e^{D_0s_2}\sigma^2(\nu-1)\frac{1}{\left(1-u^{-1}\right)^{\frac{\nu}{2}\left(1-\frac{1}{\omega}\right)}u(u-1)^{1-\nu}},
\end{equation}
we get
\begin{equation}
\int\limits_0^\infty\:  d\xi g(\xi)(\xi-z)^{-\nu}=e^{-D_0(s_1-s_2)z}.
\end{equation}
The solution $g$ is an exponential function, from which 
\begin{eqnarray}
 f(z)&=&\omega
  \frac{e^{-iqx_0}e^{-D_0s_2}}{\sigma^2\tau(\nu-1)}
 \frac{\left[D_0(s_1-s_2)\right]^{1-\nu}}{\Gamma(1-\nu)}\\\nonumber
 &&\times e^{-D_0(s_1-s_2)z}\left(1-z^{-1}\right)^{\frac{\nu}{2}\left(1-\frac{1}{\omega}\right)}z(z-1)^{1-\nu}.
 \end{eqnarray}
The complete solution for the case $\nu<1$ now reads
\begin{eqnarray}
\fl
\tilde{P}(q,s,t,x_0,D_0)&=&
e^{-iqx_0}\exp\left(-D_0\frac{s_1-s_2\rho e^{-\omega t/\tau}}{1-\rho e^{-\omega t/\tau}}\right)
\\\nonumber
\fl &&\times
\left(e^{-\omega t/\tau}\right)^{\frac{\nu}{2}\left(1-\frac{1}{\omega}\right)}\left(\frac{1-\rho e^{-\omega t/\tau}}{1-\rho}\right)^{-\nu}
\\\nonumber
\fl &&\times
\frac{1}{\Gamma(1-\nu)}\:\gamma\left(1-\nu,\frac{D_0\omega}{\sigma^2\tau}
\frac{(1-\rho)}{\left(1-\rho e^{-\omega t/\tau}\right)}\frac{e^{-\omega t/\tau}}{\left(1-e^{-\omega t/\tau}\right)}\right),
\end{eqnarray}
where $\gamma(a,x)$ is the lower incomplete gamma function $\gamma(a,x)=\int\limits_0^xdu\: e^{-u}u^{a-1}$.
The probability of $D>0$ can be obtained by integrating over $x$ and $D$ (setting respectively $q=0$ and $s=0$). As a consequence, we retrieve Feller's formula for the probability \cite{Feller1951}
\begin{equation}
\pi(D=0,t\vert D_0)=1-\frac{1}{\Gamma(1-\nu)}\gamma\left(1-\nu,\frac{D_0}{\sigma^2\tau}\frac{e^{-t/\tau}}{\left(1-e^{-t/\tau}\right)}\right).
\end{equation}
Once the process reaches the absorbing boundary at $D=0$, it remains trapped there, so that the probability  $\pi(D=0,t\vert D_0)$ is a norm decreasing function of time. 
Figure \ref{fig:prob_D_equal_zero} shows the behavior of the probability $D=0$ as a function of $\nu$. At all times, this probability is zero at $\nu=1$ and is equal to $1$ at $\nu=0$. As time increases, the diffusivity distribution is getting localized at $D=0$.

\section{Subordination}
\label{sec:Subordination}
Subordination is an elegant mathematical tool to describe complex processes, in particular anomalous diffusion \cite{Weron2009,Thiel2013}. Chechkin {\it et al.} \cite{Chechkin2016} applied it in the diffusing diffusivity context by observing that the Fokker-Planck equation
\begin{equation}\label{eq:subor_FP_D_t}
\frac{\partial}{\partial t}P(x,t) = D(t)\frac{\partial^2}{\partial x^2}P(x,t),
\end{equation}
can be written in the subordinated form:
\begin{equation}
\left\{
    \begin{array}{ll}
\frac{\partial p(x,u)}{\partial u}=\frac{\partial^2}{\partial x^2}p(x,u), \\
\frac{\partial u}{\partial t}=D(t),   \end{array}
\right.
\end{equation}
where $p(x,u)=\frac{1}{\sqrt{4\pi u}}\exp\left(-\frac{x^2}{4u}\right)$ is the Gaussian propagator. Let $T(u,t)$ be the probability density of $u(t)=\int\limits_0^tD(s)ds$. The solution of Eq. (\ref{eq:subor_FP_D_t}) can be expressed as

\begin{equation}
P(x,t)=\int\limits_0^\infty p(x,u)T(u,t)du=\int\limits_0^\infty \frac{e^{-\frac{x^2}{4u}}}{\sqrt{4\pi u}}T(u,t)du.
\end{equation}
Now the Fourier transform with respect to $x$ yields:
\begin{equation}\label{eq:char_fun_subord_integfrom}
\tilde{P}(q,t)= \int\limits_0^\infty T(u,t)e^{-q^2u}du=\tilde{T}(q^2,t),
\end{equation}
where $\tilde{T}(q^2,t)$ denotes the Laplace transform of $T$ with respect to $s=q^2$.
In our model, the description of diffusivity is made in term of the Cox-Ingersoll-Ross equation which reads
\begin{equation}
\frac{\partial \Pi(D,t\vert D_0)}{\partial t}=\frac{1}{\tau}\frac{\partial}{\partial D}\left[(D-\bar{D})\Pi\right]+\sigma^2\frac{\partial^2}{\partial D^2}\left(D\Pi\right).
\end{equation}
In the Laplace domain, one has 
\begin{equation}
\frac{\partial}{\partial t}\tilde{\Pi}(s,t)+G(s)\frac{\partial}{\partial s}\tilde{\Pi}(s,t)=-\frac{1}{\tau}\bar{D} s\tilde{\Pi}(s,t),
\end{equation}
with $G(s)=s(\sigma^2s+\frac{1}{\tau})$.
 The initial condition is now $\tilde{\Pi}(s,t=0\vert D_0)=e^{-sD_0}$. The integral $\tilde{T}(s,t)=\int\limits_0^t\tilde{\Pi}(s,t')dt'$ is known from \cite{Dufresne1990}
\begin{eqnarray}\nonumber
T(s,t|D_0)&=&\left[\frac{e^{t^*/2}}{\cosh(\omega_s t^*/2)+\frac{1}{\omega_s}\sinh(\omega_s t^*/2)}\right]^{\nu}\\
&&\times\exp\left[-\frac{sD_0\tau}{\omega_s}
\frac{2\sinh(\omega_s t^*/2)}
{\cosh(\omega_s t^*/2)+\frac{1}{\omega_s}\sinh(\omega_s t^*/2)}\right],
\end{eqnarray}
with $t^*=t/\tau$ and $\omega_s=\sqrt{1+4s\sigma^2\tau^2}$. According to Eq. (\ref{eq:char_fun_subord_integfrom}), one deduces thus the characteristic function as a function of initial diffusivity $D_0$:
\begin{eqnarray}\nonumber
\tilde{P}(q,t|D_0)&=&\left[\frac{e^{t^*/2}}{\cosh(\omega t^*/2)+\frac{1}{\omega}\sinh(\omega t^*/2)}\right]^{\nu}\\
&&\times\exp\left[-\frac{D_0q^2\tau}{\omega}
\frac{2\sinh(\omega t^*/2)}
{\cosh(\omega t^*/2)+\frac{1}{\omega}\sinh(\omega t^*/2)}\right],
\end{eqnarray}
with $\omega=\sqrt{1+4q^2\sigma^2\tau^2}$.
After integration over initial diffusivity the characteristic function yields
\begin{eqnarray}\nonumber
\tilde{P}(q,t)&=&\left[\frac{e^{t^*/2}}{\cosh(\omega t^*/2)+\frac{1}{\omega}\sinh(\omega t^*/2)}\right]^{\nu}
\\
&&\times\left(1+\frac{2\sigma^2q^2\tau^2
\sinh(\omega t^*/2)}{\omega\cosh(\omega t^*/2)+\left(1-2\omega\sigma^2q^2\tau^2\right)\sinh(\omega t^*/2)
}\right)^\nu.
\end{eqnarray}
This is an alternative representation of the characteristic function $\tilde{P}(q,t)$ from Eq. (\ref{eq:marginal_solution}).

\section{Autocorrelation of squared increments}
\label{sec:Deriv_autoc_squared_inc}
We have $dx_t=\sqrt{2D_t}dW_t^{(1)}$ and the diffusivity in the integral form reads
\begin{equation}\label{eq:diffusivity_int_form}
D_t=D_0e^{-t/\tau}+\bar{D} (1-e^{-t/\tau})+e^{-t/\tau}\int_0^te^{s/\tau}\sqrt{D_s}dW_s^{(2)}.
\end{equation}
We define the centered squared increments $dx_t^{2*}=dx_t^2-\langle dx_t^2\rangle$.
\\
Their autocorrelation is then
\begin{eqnarray}
\fl
\langle dx_t^{2*}dx_{t+\Delta}^{2*}\rangle &=& \langle dx_t^2dx_{t+\Delta}^2 \rangle-\langle dx_t^2\rangle \langle dx_{t\Delta}^2 \rangle\\\nonumber
&=&4\langle D_tD_{t+\Delta}\rangle\left\langle \left(dW_t^{(1)}dW_{t+\Delta}^{(1)}\right)^2\right\rangle\\
&& - 4\langle D_t\rangle\langle D_{t+\Delta}\rangle\left\langle \left(dW_t^{(1)}\right)^2\right\rangle\left\langle \left(dW_{t+\Delta}^{(1)}\right)^2\right\rangle .
\end{eqnarray}
For $\Delta=0$, we calculate
\begin{equation}
\langle (dx_t^{2*})^2\rangle=12\langle D_t^2\rangle- 4\langle D_t\rangle^2,
\end{equation}
which is obtained directly from Eq. (\ref{eq:diffusivity_int_form}):
\begin{eqnarray}\nonumber
\langle (dx_t^{2*})^2\rangle= && 12\left[\sigma^2\bar{D}\tau\left(1-e^{-t/\tau}\right)^2+2\sigma^2\tau D_0\left(e^{-t/\tau}-e^{-2t/\tau}\right)\right]\\
&&+
\quad 8\left(D_0e^{-t/\tau}+\bar{D}\left(1-e^{-t/\tau}\right)\right)^2.
\end{eqnarray}
In the case $\Delta>0$, as $dW_t^{(1)}$ is independent from $dW_{t+\Delta}^{(1)}$, one has $\left\langle \left(dW_t^{(1)} dW_{t+\Delta}^{(1)}\right)^2\right\rangle=\left\langle\left(dW_t^{(1)}\right)^2 \right\rangle\left\langle \left(dW_{t+\Delta}^{(1)}\right)^2\right\rangle$ which leads to

\begin{equation}
\langle dx_t^{2*}dx_{t+\Delta}^{2*}\rangle=
4\langle D_tD_{t+\Delta}\rangle- 4\langle D_t\rangle\langle D_{t+\Delta}\rangle.
\end{equation}
The autocorrelation of squared increments is explicitly related to the autocorrelation of diffusivity as
\begin{equation}
\fl
\langle dx_t^{2*}dx_{t+\Delta}^{2*}\rangle=4e^{-(2t+\Delta)/\tau}
\int_0^t\int_0^{t+\Delta} e^{(s_1+s_2)/\tau}\langle\sqrt{D_{s_1}D_{s_2}}\rangle\langle dW_{s_1}^{(2)}dW_{s_2}^{(2)}\rangle,
\end{equation}
from which
\begin{equation}
\fl
\langle dx_t^{2*}dx_{t+\Delta}^{2*}\rangle=4
e^{-\Delta/\tau}\left[\sigma^2\bar{D}\tau\left(1-e^{-t/\tau}\right)^2+2\sigma^2\tau D_0\left(e^{-t/\tau}-e^{-2t/\tau}\right)\right].
\end{equation}
\section{Asymptotic analysis at large $x$}
\label{Annexe:Asymp_large_x}
We investigate the asymptotic behavior of the propagator at large $x$.
Let us first consider the particular case $\nu = 1$.  As the
propagator $P(x,t|x_0)$ is obtained as the inverse Fourier transform
of $\tilde{P}(q,t)$, it is instructive to search for the poles of
$\tilde{P}(q,t)$ in the complex plane of $q$ in order to compute the
inverse Fourier transform by the residue theorem.  We write
\begin{equation}
\tilde{P}(q,t) = \frac{\omega \, e^{t^*/2} }{f_{+}(\omega) \, f_{-}(\omega)},
\end{equation}
where
\begin{eqnarray}
f_+(\omega) & = &\omega \cosh (t^*\omega/4) + \sinh(t^*\omega/4)  , \\
f_-(\omega) & = &\omega \sinh (t^*\omega/4) + \cosh(t^*\omega/4)  . 
\end{eqnarray}
Setting $\omega = i 4\alpha/t^*$, we search for $\alpha$
at which these functions vanish, i.e.,
\begin{eqnarray}
f_+(\omega) & = i(4\alpha/t^*) \cos(\alpha) + \sin(\alpha) = 0 , \\
f_-(\omega) & = - (4\alpha/t^*) \sin(\alpha) + \cos(\alpha) = 0 . 
\end{eqnarray}
Both equations have infinitely many solutions.  It is easy to see that
the solutions of the first equation lie in the intervals
$\bigcup\limits_{k=-\infty}^\infty (\pi/2 + k\pi,\pi + k\pi)$
(including the trivial solution $\alpha = 0$), whereas the solutions
of the second equation lie in the intervals
$\bigcup\limits_{k=-\infty}^\infty (k\pi,\pi/2 + k\pi)$.  Since
$\omega = 0$ is not a pole of $\tilde{P}(q,t)$ (as it is compensated by the
numerator), we exclude this point.  The pole with the smallest
absolute value is thus given as the smallest positive solution of the
second equation that we rewrite as
\begin{equation}
\alpha_{t^*} \sin \alpha_{t^*} = \frac{t^*}{4} \cos\alpha_{t^*}  .
\end{equation}
The smallest positive solution of this equation, $\alpha_{t^*}$, is a
monotonously increasing function of $t^*$, ranging from $0$ at $t^*
= 0$ to $\pi/2$ at $t^* = \infty$.  The corresponding value of
$\omega$ will determine the asymptotic exponential decay of the
propagator.

Since $i 4\alpha_{t^*}/t^* = \omega = \sqrt{1 + 4q^2 \sigma^2\tau^2}$,
we identify the pole in the $q$ plane:
\begin{equation}
q_0 = \pm i \beta_{t^*}  \frac{1}{2\sigma\tau}  \,, \qquad \beta_{t^*} = \sqrt{1 + (4\alpha_{t^*}/t^*)^2} \,.
\end{equation}
Applying the residue theorem, we get
\begin{equation}
P(x,t|x_0) = \int\limits_{-\infty}^\infty \frac{dq}{2\pi} e^{iq(x-x_0)} \tilde{P}(q,t)
= 2\pi i \sum\limits_n \frac{e^{iq_n(x-x_0)}}{2\pi} \res_{q_n} \{\tilde{P}(q,t)\},
\end{equation}
where the sum runs over the poles. The asymptotic behavior at large $|x-x_0|$ is determined by the pole with the smallest $|q_0|$. We get thus Eq. (\ref{asymp_large_x_nu1}). One can also compute the prefactor by evaluating the residue of $\tilde{P}(q,t)$ at $q = q_0$. Note that for large $t^*$, one has $\alpha_{t^*} \approx \pi/2$, and
thus the dependence on $t^*$ is eliminated, yielding $\beta_{t^*}
\simeq 1$ as $t^*\to\infty$.  In turn, when $t^*$ is small, one has
$\alpha_t^* \simeq \sqrt{t^*}/2$, and thus $\beta_{t^*} \simeq
\sqrt{1 + 4/t^*} \to \infty$.  As a consequence, the distribution becomes more and more narrowed, as expected. We emphasize that this analysis is not rigorous enough, as the relation between $q$ and $\omega$ involves the square root and thus requires some cuts in the complex plane to avoid multiple branches.

When $\nu$ is a strictly positive integer, the above analysis remains
applicable.  However, the pole is not simple (as for $\nu = 1$) but
has a degree $\nu$.  The degree $\nu > 1$ results in a more
complicated computation of the residue and, more importantly, in power law corrections to the exponential decay in Eq. (\ref{asymp_large_x_nudiff1}).

We also emphasize that the current analysis only focuses on the dependence on $|x-x_0|$ and does not capture the complete dependence on $t^*$ which enters through different coefficients.



\vskip 5mm
\begin{thebibliography}{98}


\bibitem{Weiss2004}   Weiss M, Elsner M,  Kartberg F, and Nilsson T, 2004,  
{\it Biophys. J.}, {\bf 87}, 3518–24.

\bibitem{Barkai2012} Barkai E, Garini Y, and  Metzler R, 2012,
 {\it Phys. Today}, {\bf 65}, 29–35.

\bibitem{Bertseva2012} Bertseva E, Grebenkov D S, Schmidhauser P, Gribkova S, Jeney S and Forro L, 2012, {\it Eur. Phys. J. E}, {\bf 35}, 63.

\bibitem{Hofling2013} H\"{o}fling F and  Franosch T, 2013,
{\it Rep. Prog. Phys.}, {\bf 76}, 046602.

\bibitem{Manzo2015} Manzo C, Torreno-Pina J A, Massignan P, Lapeyre G J, Lewenstein M, and Garcia Parajo M F, 2015 
{\it Phys. Rev. X}, {\bf 5}, 14.

\bibitem{Sadegh2017} Sadegh S, Higgins J L, Mannion P C, Tamkun M M, and Krapf D, 2017,
{\it Phys. Rev. X}, {\bf 7}, 11031.

\bibitem{Bouchaud1990} Bouchaud J-P and Georges A 1990, {\it Phys. Rep.}, {\bf 195}, 127.                      
 

\bibitem{Havlin2002} Havlin S and Ben-Avraham D, 2002, 
{\it Adv. Phys.}, {\bf 36}, 187-292.


\bibitem{Metzler2014} Metzler R, Jeon J H, Cherstvy A G, and Barkai E, 2014,
{\it Phys. Chem. Chem. Phys.}, {\bf 16}, 24128-24164.

\bibitem{Spanner2016} Spanner M, H\"{o}fling F, Kapfer S C, Mecke K R, Schr\"{o}der-Turk G E, and Franosch T, 2016,
{\it Phys. Rev. Lett.}, {\bf 116}, 060601.

\bibitem{Mandelbrot1968}
Mandelbrot B B, Van Ness J W, 1968,
{\it SIAM Rev.}, {\bf 10}, 4.


\bibitem{Wang1992} Wang K G, 1992, {\it Phys.Rev. A}, {\bf 45}, 833.

\bibitem{Porra1996} Porr\'a J M, Wang K G, Masoliver J, 1996, {\it Phys.Rev. E}, {\bf 53}, 5872.

\bibitem{Grebenkov2011b} Grebenkov D G, 2011, {\it Phys.Rev. E} {\bf 83}, 061117.

\bibitem{Montroll1965}
Montroll E W and Weiss G H, 1965,
{\it J. Math. Phys.}, {\bf 6}, 167.

\bibitem{Metzler2000} Metzler R and Klafter J, 2000, {\it Phys. Rep.}, {\bf 339}, 1-77.


\bibitem{Orpe2007}	Orpe A V and Kudrolli A, 2007, 
{\it Phys. Rev. Lett.}, {\bf 98}, 238001.

\bibitem{Gollub1991}	Gollub J P, Clarke J, Gharib M, Lane B, and Mesquita O N, 1991, 
{\it Phys. Rev. Lett.}, {\bf 67}, 3507-3510.

\bibitem{Stuhrmann2012}	Stuhrmann B, Soares E Silva M, Depken M, MacKintosh F C and Koenderink G H, 2012,
{\it Phys. Rev. E}, {\bf 86}, 1–5.

\bibitem{Toyota2011}	Toyota T, Head D A, Schmidt C F and Mizuno D, 2011, 
{\it Soft Matter}, {\bf 7}, 3234.


\bibitem{Bertrand2012}	Bertrand O J N, Fygenson D K and Saleh O A, 2012,
{\it Proc. Natl. Acad. Sci. USA}, {\bf 109}, 17342-17347.

\bibitem{Moschakis2012}	Moschakis T, Lazaridou A and Biliaderis C G, 2012, 
{\it J Coll. Inter. Sci.}, {\bf 375}, 50–59.

\bibitem{Chaudhuri2007}	Chaudhuri P, Berthier L and Kob W, 2007, 
{\it Phys. Rev. Lett.}, {\bf 99}, 060604.


\bibitem{Grady2017}	Grady M E, Parrish E, Caporizzo M A, Seeger S C, Composto R J, Eckmann D M, 2017, 
{\it Soft Matter}, {\bf 13}, 1873-1880.

\bibitem{Wang2009}	Wang B, Anthony S M, Bae S C and Granick S, 2009, 
{\it Proc. Nat. Acad. Sci.}, {\bf 106}, 15160-15164.

\bibitem{Wang2012}	Wang B, Kuo J, Bae S C and Granick S, 2012 
{\it Nat. Mater.} {\bf 11}, 481-485.

\bibitem{Dragulescu2011} Dr\v{a}gulescu A A and Yakovenko V M, 2011, 
{\it Quant. Fin.}, {\bf 2}, 443-453.

\bibitem{Rouyer2000} Rouyer F and Menon N, 2000, 
{\it Phys. Rev. Lett.}, {\bf 85}, 3676-3679.

\bibitem{He2016}	He W, Song H, Su Y, Geng L, Ackerson B J, Peng H B and Tong P, 2016, 
{\it Nat Com.}, {\bf 7}, 11701.

\bibitem{Ghosh2015} Ghosh S K, Cherstvy A G and Metzler R, 2015, 
{\it Phys. Chem. Chem. Phys.}, {\bf 17} 1847.

\bibitem{Ghosh2016} Ghosh S K, Cherstvy A G, Grebenkov D S and Metzler R, 2016, 
{\it New. J. Phys.}, {\bf 18}, 013027.

\bibitem{Sarnanta2016} Sarnanta N and Chakrabarti R, 2016, {\it Soft Matter}, {\bf 12}, 8554.


\bibitem{Metzler2017a} Metzler R, 2017, 
{\it Biophys. J.}, {\bf 112}, 413-447.

\bibitem{Karger1985}
K\"{a}rger J, 1985, 
{\it Adv. Coll. Int. Sci.}, {\bf 23}, 129-148.

\bibitem{Fieremans2010} Fieremans E, Novikov D S, Jensen J H, Helpern J A, 2010, 
{\it NMR Biomed. {\bf 23}, 711-724}.

\bibitem{Chubynsky2014} Chubynsky M V and Slater G W, 2014, 
{\it Phys. Rev. Lett.}, {\bf 113}, 098302.

\bibitem{Beck2003} Beck C and Cohen E G D, 2003, 
{\it Physica A}, {\bf 322}, 267-275.

\bibitem{Beck2005} Beck C, Cohen E G D, and Swinney HL, 
{\it Phys. Rev. E}, {\bf 72}, 056133.

\bibitem{Jain2016}
Jain R and Sebastian K L, 2016, 
{\it J. Phys. Chem. B}, {\bf 120}, 3988-3992.

\bibitem{Jain2017b} Jain R and Sebastian K L, 2017, 
{\it J. Chem. Sci.}, {\bf 126}, 929-937.

\bibitem{Jain2017a} Jain R and Sebastian K L, 2017, 
{\it Phys. Rev. E}, {\bf 95}, 032135.

\bibitem{Chechkin2016} Chechkin A V, Seno F, Metzler R, and Sokolov I M, 2017, 
{\it Phys. Rev. X}, {\bf 7}, 021002.

\bibitem{Higham2001}  Higham D J, 2001, 
{\it SIAM Rev.}, {\bf 43}, 525-546.

\bibitem{Feller1951} Feller W, 1951, 
{\it Ann. Math.} {\bf 54}, 173-182.

\bibitem{Gan2015}
Gan X, Waxman D, 2015, {\it Phys .Rev. E}, {\bf 91}, 012123.

\bibitem{Cox1985}
Cox J C, Ingersoll J E, Ross S A, 1985, 
{\it Econometrica}, {\bf 53}, 385-408.

\bibitem{Heston1993}
Heston S L, 1993, 
{\it Rev. Fin. Studies.}, {\bf 6}, 327-343.

\bibitem{Chen1996}
Chen L, 1996, 
{\it Financial Markets, Institutions and Instruments.}, {\bf 5}, 1-88.

\bibitem{He2008}
He Y, Burov S, Metzler R and Barkai E, 2008, {\it Phys. Rev. Lett.}, {\bf 101}, 058101.

\bibitem{Burov2010}
Burov S, Jeon J H, Metzler R, and Barkai E, 2010, 
{\it Phys. Chem. Chem. Phys.}, {\bf 13}, 1800-1812.

\bibitem{Metzler2017b} Schwarzl M, Godec A, and Metzler R, 2017, 
{\it Sci. Rep.}, {\bf 7}, 3878.

\bibitem{Grebenkov2011} Grebenkov D S, 2011, 
{\it Phys. Rev. E}, {\bf 84}, 031124.


\bibitem{Sikora2017} Sikora G, Teuerle M, Wyłomańska A, and Grebenkov D S, 2017, 
{\it Phys. Rev. E}, {\bf 96}, 022132.

\bibitem{Qian1991} Qian H, Sheetz M P, Elson E L, 1991, {\it Biophys. J.}, {\bf 60}, 910-921. 

\bibitem{Cherstvy2016} Cherstvy A G and Metzler R, 2016, 
{\it Phys. Chem. Chem. Phys.}, {\bf 18}, 23840-23852.

\bibitem{Kindermann2016} Kindermann F, Dechant A, Hohmann M, Lausch T, Mayer D, Schmidt F, Lutz E and Widera L, 2016, 
{\it Nat. Phys.}, {\bf 13}, 137-141.

\bibitem{Magdziarz2011}
Magdziarz M and Weron A, 2011, 
{\it Phys. Rev. E}, {\bf 84}, 051138.

\bibitem{Lanoiselee2016}
Lanoisel\'ee Y and Grebenkov D S, 2016, 
{\it Phys. Rev. E}, {\bf 93}, 052146.

\bibitem{Jensen2005} Jensen J H, Helpern J A, Ramani A, Lu H, Kaczynski K, 2005, 
{\it Magn. Reson. Med.}, {\bf 53}, 1432-1440.

\bibitem{Valle1991}
Valle A, Rodriguez M A and Pesquerra L, 1991, {\it Phys. Rev. A}, {\bf 43}, 948.


\bibitem{McCauley2013} McCauley J J, 2013, 
{\it Cambridge University Press}.


\bibitem{Weron2009}
Weron A and Magdziarz M, 2009, 
{\it Eur. Phys. Lett.}, {\bf 86}, 60010.

\bibitem{Thiel2013} Thiel F, Flegel F, and Sokolov I M, 2013, 
{\it Phys. Rev. Lett.}, {\bf 111}, 010601.

\bibitem{Dufresne1990} 
Dufresne D, 2001, 
{\it Working Paper, University of Melbourne}.

\bibitem{Hernandez1990a} Hern\'{a}ndez-Garc\'{i}a E, Rodr\'{i}guez M A, C\'{a}ceres M O and San Miguel M, 1990, {\it Phys. Rev. A}, {\bf 41}, 4562.
\bibitem{Hernandez1990b} Hern\'{a}ndez-Garc\'{i}a E and C\'{a}ceres M O, 1990, {\it Phys. Rev. A}, {\bf 42}, 4503.
\end{thebibliography}
\end{document}